\documentclass[a4paper,11pt]{article}
\usepackage{jheppub} 
\usepackage{lineno}
\usepackage{amsmath,amsthm,amssymb}
\usepackage{minibox,graphbox}
\usepackage{graphicx}
\usepackage{physics}
\usepackage{slashed}
\usepackage{mathtools}                      
\usepackage{bm}  
\usepackage{comment}
\usepackage{subcaption}
\newcommand{\Ai}{\operatorname{Ai}}

\newcommand{\ngin}{n_{\gamma,\text{in}}}
\newcommand{\ngout}{n_{\gamma,\text{out}}}

\newcommand{\Wtr}{W_{\rm dir, 2-loop}}

\newcommand{\mM}{\mathcal{M}}
\newcommand{\mP}{\mathcal{P}}
\newcommand{\mT}{\mathcal{T}}
\newcommand{\mF}{\mathcal{F}}

\title{\boldmath Cutting rules in strong field QED with application to trident pair production}

\author[a]{Y. V. Selivanov}
\author[b]{A. A. Mironov}
\author[a]{A. I. Alexeenko}
\author[a]{A. M. Fedotov}
\affiliation[a]{National Research Nuclear University MEPhI, Moscow, 115409, Russia}
\affiliation[b]{CPHT, CNRS, \'Ecole polytechnique, Institut Polytechnique de Paris, 91128 Palaiseau, France}

\emailAdd{mironov.hep@gmail.com}

\abstract{Following Veltman’s approach, we formulate and discuss a general cutting equation for QED in a plane-wave background. We apply the corresponding cutting rules to justify the connection between the two-loop radiative corrections to elastic electron scattering and the rate of the trident process in a constant crossed field. Among other terms generated by cuttings, some field-induced radiation corrections also emerge, which contain UV-finite contributions from loop insertions into external legs. As a byproduct, we compare the previously published results for the trident process in a constant crossed field and present a complete analytical expression for direct and exchange contributions to its rate, which is resolved in the spin of the initial electron. Our findings establish that although total rates can be reliably extracted from higher-loop by applying the cutting rules, reconstruction of differential rates requires additional care. The cutting rules apply to any loop order and may be extended to nonperturbative regimes.}

\begin{document}
\maketitle
\flushbottom

\section{Introduction}
Cutting rules for scattering amplitudes in quantum field theories (QFTs) appear among the vital theoretical tools. They are connected to unitarity \cite{cutkosky1960singularities, veltman1963unitarity} and are practical for extracting decay rates and dispersion relations \cite{cutkosky1960some, ball1962scattering}. Unitarity cuts have become a basis for computational techniques in particle physics \cite{bern1994one,bern2007shell,ossola2008cuttools, ellis2012one, cascioli2012scattering}. 

For QFTs beyond textbook cases \cite{peskin2018introduction, dixon2014brief, zwicky2016brief}, the cutting rules may become no more trivial and might need reformulation. Among such nontrivial examples are theories in thermal background, \cite{carrington2003scattering}, consideration of amplitudes in superstring or conformal field theories \cite{pius2016cutkosky, meltzer2020cft}, and applications in cosmology \cite{melville2021cosmological}.

Another nontrivial extension is adding a strong classical background. In strong-field quantum electrodynamics (SFQED), the interaction can be altered, as particles get coupled to the external field \cite{greiner2005quantum, fedotov2023high}. Depending on the field structure, some theory backbones may require reconsideration, e.g. ways of building the perturbation theory  \cite{hernandez2023strong}, the infrared behavior \cite{klisch2026leading}, or how observables should be defined \cite{edwards2021resummation}. The optical theorem formulation falls into this category as well, and, for example, has been revised for QED in pair-creating fields \cite{fradkin1988optical, fradkin1991book}. In this work, we will focus on a plane-wave background and a constant crossed field (CCF; in this configuration, electric and magnetic components are equal and perpendicular, $E=H$, $\vec{E}\perp\vec{H}$), which can be viewed as the limiting case of a plane wave at zero frequency. Such fields are null and hence do not create pairs spontaneously. These field models often apply to laser-matter interactions at extreme intensities \cite{Di_Piazza_2012, Gonoskov_2022}. Furthermore, within the locally constant field approximation (LCFA), for ultrarelativistic particles, a CCF is a universal background model, as it covers a wide class of adiabatically varying space- and time-dependent field configurations produced by any type of source \cite{fedotov2023high}. The symmetry of a CCF appears particularly convenient, as it allows additional simplifications when considering scattering amplitudes.

The nonperturbative coupling of particles to the background is usually treated within the strong-field approach \cite{furry1951bound, sokolov1952quantum, keldysh1958effect}. Such coupling may induce a range of new or modify existing effects \cite{fedotov2023high, popruzhenko2023dynamics}, which scale both with the particle energy and the background strength \cite{erber1966high, baier1972higher, ritus1985quantum}. The kinematically allowed processes include $1\rightarrow n$ transitions  \cite{nikishov1964quantum}. Loop corrections are also modified, e.g. in constant electromagnetic fields \cite{narozhny1969propagation, ritus1972radiative, shabad1975photon}, and are expected to reveal in observable effects \cite{meuren2011quantum, podszus2021first, akhmedov2023loop, li2023strong}. The extraordinary feature of higher-order loop contributions is their rapid growth with the field strength \cite{narozhny1979radiation,narozhny1980expansion,fedotov2017conjecture}, leading to essential nonperturbativity and requiring resummations \cite{gusynin1999electron, mironov2020resummation}. 

Application of the optical theorem in SFQED is tempting for practical calculations of scattering processes beyond leading order, as some higher-order loop results are available in the literature (partially covered below). A direct computation of matrix elements even at the second order of perturbation theory is challenging due to technical difficulties \cite{fedotov2023high, seipt2012two, mackenroth2013nonlinear, dinu2018trident}. At the same time, such calculations have become in high demand recently due to unprecedented advancement in experimental capabilities \cite{Gonoskov_2022, bamber1999studies, abramowicz2021conceptual, nielsen2023precision, nielsen2023differential, storey2023status}. At one loop, the unitarity cuts are straightforward and have proven robust: in a magnetic field \cite{tsai1974photon, shabad1975photon}, a CCF \cite{ritus1972radiative}, a plane wave \cite{baier1976interaction, baier1976theory, meuren2013polarization}, and even in laser pulses \cite{meuren2015polarization}. In a CCF, the cutting method was also applied at two loops to extract the rates of the trident process (creation of a pair by an electron via a virtual photon) \cite{ritus1972vacuum}, two-photon emission by an electron \cite{morozov1975elastic}, and photo-trident process \cite{morozov1977elastic}. Note that in many cases, the approximation of splitting a higher-order diagram into a chain of first-order processes is common and robust in a strong field limit \cite{kirk2008, sokolov2010pair, tang2023locally, king2024feasibility, pouyez2025kinetic, blackburn2026revealing}, however, it neglects the coherence and interference features that may be important in the intermediate regimes \cite{king2018effect, mackenroth2018nonlinear, torgrimsson2020nonlinear}.

Although cutting rules in SFQED have already found a handful of applications, their general formulation is overlooked. In works~\cite{ritus1972vacuum,morozov1975elastic,morozov1977elastic}, a correct cutting operation was introduced ad hoc (for a CCF); however, it was not fully substantiated. Furthermore, the comparison of the results from the direct matrix element evaluation and the imaginary part of the loop amplitudes is uncommon beyond leading order. Derivation of such rules in a general plane-wave background is one of the main objectives of this work. We build our treatment on a general approach developed by Veltman \cite{veltman1963unitarity,veltman1994diagrammatica}. It introduces the so-called largest time equation (LTE) for general Feynman graphs, from which the unitary cutting equation and cutting rules follow almost automatically. We prefer this powerful, though lengthy method, as it builds a consistent foundation for the analysis of QFTs with nontrivial properties, e.g. with unstable states present in the particle spectrum \cite{denner2015complex,donoghue2019unitarity}.

As the second goal of this work, we consider the cutting rules in application to the trident process. The latter appears to be one of the most well-studied tree-level second-order effects in SFQED, with results known in a CCF \cite{ritus1972vacuum, baier1972higher,  king2013trident, king2018effect}, and plane-wave backgrounds (largely in the context of intense laser fields) \cite{ilderton2011trident, hu2010complete, hernandez2019laser, krajewska2015circular, mackenroth2018nonlinear, dinu2018trident, dinu2020trident, torgrimsson2020nonlinear, dinu2020approximating, roshchupkin2025quantum, tang2025entanglement}. In the current work, we will focus on the CCF case.

The trident process amplitude splits into three terms \cite{king2013trident}: direct, exchange (in which the two final electrons are interchanged), and interference between the two.\footnote{The notation for this splitting varies in the literature. We stick to terminology used in refs.~\cite{king2013trident, king2018effect}, however, the term ``exchange'' is sometimes used for the interference term, see e.g. refs.~\cite{ritus1972vacuum, torgrimsson2020nonlinear}.} The total rate corresponding to the direct term mod-squared was evaluated from the two-loop electron elastic scattering matrix by Ritus \cite{ritus1972vacuum}. Later, the differential and total rates in their full form, and hence the direct term, were computed from the trident matrix element definition (note that we do not consider the interference contribution in the current work) \cite{king2013trident, dinu2018trident, torgrimsson2020nonlinear}. To our knowledge, the two approaches have never been consistently compared. This is likely due to differences in notations and tedious computations, resulting in lengthy expressions that are hard to comprehend, not to mention showing the direct transformation linking the two approaches. In the current work, we reconsider the trident process with both approaches within a single framework to cover this gap. We provide a derivation of the direct contribution to the trident process rate, which is fully detailed in our \texttt{Wolfram Mathematica} notebook available online \cite{github}.

The work is organized as follows. We start our discussion in section~\ref{sec:ii} by recovering the cutting rules in standard QED with Veltman's approach. In section~\ref{sec:iii}, we introduce the strong-field method and formulate the cutting equation and cutting rules for QED in a strong plane-wave background within the Furry picture. In section~\ref{sec:iv}, we illustrate our general considerations of the cutting equation with the electron elastic scattering amplitude at two loops in a general plane-wave background. Then, in section~\ref{sec:v}, we consider the polarization correction to this process in a CCF in more detail. First, we recover the explicit expression for the amplitude from the bubble-chain-resummed result \cite{mironov2020resummation}, then we discuss how the cutting approach of Ritus applies to this expression and show that it matches our cutting rules. The section is concluded by a direct check that the cutting rules indeed provide the imaginary part of the amplitude.
Section~\ref{sec:vi} is devoted to the explicit calculation of the trident process differential and total rates per the scattering matrix definition. We provide a detailed description of each step and present an explicit and comprehensible formula. Then, we calculate the total rate of the process based on the two-loop scattering matrix, and at the end, compare the two results. In section~\ref{sec:vii}, we rewrite the trident process differential rate explicitly as a function of the relative particle energies (or $\chi$ parameters, see the section), plot some examples of distributions, and discuss their behavior. Finally, we summarise and conclude our work in section~\ref{sec:viii}.

\section{Cutting equation in QED}
\label{sec:ii}
In this section, we follow Veltman's approach \cite{veltman1994diagrammatica} to derive cutting rules for QED as a starting point for further generalization to QED in a background field. The QED Lagrangian in covariant gauge is given by \cite{peskin2018introduction}\footnote{We use such units that $\hbar = c = 1$ and the metric signature $(+,-,-,-)$.}
\begin{equation}
    \mathcal{L} = -\frac{1}{4}F_{\mu\nu}F^{\mu\nu} - \frac{1}{2\xi}(\partial_\mu A^\mu)^2 + \bar{\psi}(i\slashed{\partial} - m)\psi - e\bar{\psi}\slashed{A}\psi.
\end{equation}
Free spin-$\frac{1}{2}$ field operators read
\begin{equation}\label{eq:spinor_field_free_operators}\begin{split}
    \psi(x) = \int \frac{d^3\bm{p}}{(2\pi)^3} \frac{1}{\sqrt{2\varepsilon_{\bm{p}}}} \sum_{\sigma=1}^2 \left(a_{\bm{p},\sigma}u(\bm{p}, \sigma)e^{-ipx} + b_{\bm{p},\sigma}^\dagger v(\bm{p}, \sigma)e^{ipx}\right), \\
    \bar{\psi}(x) = \int \frac{d^3\bm{p}}{(2\pi)^3} \frac{1}{\sqrt{2\varepsilon_{\bm{p}}}} \sum_{\sigma=1}^2 \left(b_{\bm{p},\sigma} \bar{v}(\bm{p}, \sigma)e^{-ipx} + a_{\bm{p},\sigma}^\dagger \bar{u}(\bm{p}, \sigma)e^{ipx}\right),
    \end{split}
\end{equation}
where $p^0 = \varepsilon_{\bm{p}} = \sqrt{\bm{p}^2+m^2}$. These operators obey the canonical anticommutation relations. Bispinor amplitudes $u_\sigma(\bm{p},\sigma)$ and $v_\sigma(\bm{p},\sigma)$ satisfy the following conditions:
\begin{equation}\label{eq:u_v_properties}
    \begin{split}(\slashed{p}-m)u(\bm{p,\sigma}) = 0, \quad (\slashed{p}+m)v(\bm{p},\sigma) = 0, \\
    \bar{u}(\bm{p}, \sigma)u(\bm{p}, \sigma') = 2m \delta_{\sigma\sigma'}, \quad \bar{v}(\bm{p}, \sigma)v(\bm{p}, \sigma') = -2m \delta_{\sigma\sigma'}, \\
    \sum_\sigma u(\bm{p}, \sigma)\bar{u}(\bm{p}, \sigma) = \slashed{p}+m, \quad \sum_\sigma v(\bm{p}, \sigma)\bar{v}(\bm{p}, \sigma) = \slashed{p}-m.
    \end{split}
\end{equation}

Now, consider the Feynman propagator
\begin{equation}
    \begin{split}
    S_{ab}(x - y) \equiv & \langle 0|T\psi_a(x)\bar{\psi}_b(y)|0\rangle \\ = & \,\theta(x^0-y^0)\bra{0}\psi_a(x)\bar{\psi}_b(y)\ket{0} - \theta(y^0-x^0)\bra{0}\bar{\psi}_b(y)\psi_a(x)\ket{0} \\
    \equiv & \,\theta(x^0-y^0)S_{ab}^{(+)}(x-y) + \theta(y^0-x^0)S_{ab}^{(-)}(x-y),
    \end{split}
\end{equation}
where $S^{(+)}$ and $S^{(-)}$ are the positive- and negative-energy Wightman functions, respectively, given by
\begin{equation}
    \begin{split}\label{eq:Wightman_spinor_field}
    S^{(+)}(x-y) & = \int \frac{d^3 \bm{p}}{(2\pi)^3} \frac{1}{2\varepsilon_{\bm{p}}}\left.\sum_\sigma u(\bm{p}, \sigma)\bar{u}(\bm{p}, \sigma)e^{-ip(x-y)}\right|_{p^0=\varepsilon_{\bm{p}}} \\
    & = \int \frac{d^4 p}{(2\pi)^3}\theta(p^0)\delta(p^2-m^2)(\slashed{p}+m)e^{-ip(x-y)}, \\
    S^{(-)}(x-y) & = -\int \frac{d^3 \bm{p}}{(2\pi)^3} \frac{1}{2\varepsilon_{\bm{p}}}\left.\sum_\sigma v(\bm{p}, \sigma)\bar{v}(\bm{p}, \sigma)e^{ip(x-y)}\right|_{p^0=\varepsilon_{\bm{p}}} \\
    & = \int \frac{d^4 p}{(2\pi)^3}\theta(-p^0)\delta(p^2-m^2)(\slashed{p}+m)e^{-ip(x-y)}.
    \end{split}
\end{equation}
Note that the change of integration variable $p \rightarrow-p$ was made in the last step. It is easy to see that the Wightman functions for the free spin-$\frac{1}{2}$ field satisfy the condition
\begin{equation}\label{eq:conjugate_fermion_propagator}
    \bar{S}^{(\pm)}(y-x) \equiv \gamma^0 \left(S^{(\pm)}(y-x)\right)^\dagger \gamma^0 = S^{(\pm)}(x-y).
\end{equation}
Thus, we can write  
\begin{equation}\label{eq:spinor_field_free_separation}
    \begin{split} S(x-y) = \theta(x^0-y^0)S^{(+)}(x-y) + \theta(y^0-x^0)S^{(-)}(x-y), \\
    \bar{S}(y-x) = \theta(x^0-y^0)S^{(-)}(x-y) + \theta(y^0-x^0)S^{(+)}(x-y).
  \end{split}
\end{equation}
The latter relations will be key to formulating the cutting rules. 

Similar relations can be obtained for the electromagnetic field, for which the Feynman propagator in the $R_\xi$-gauge is given by
\begin{equation}\label{eq:em_field_feynman_propagator}
    D_{\mu\nu}(x-y) = \int \frac{d^4 l}{(2\pi)^4}\frac{-i(g_{\mu\nu} - (1-\xi)l_\mu l_\nu /l^2)}{l^2+i0}e^{-il(x-y)}.
\end{equation}
Due to gauge invariance, for QED it is sufficient to derive the cutting rules in the Feynman gauge with $\xi = 1$. For the spin sum, the Ward-Takahashi identity justifies the substitution
\begin{equation}\label{eq:WTidentity}
    \sum_\lambda \epsilon_\mu(\bm{l},\lambda)\epsilon_\nu^*(\bm{l},\lambda) \rightarrow -g_{\mu\nu}.
\end{equation}
Using the residue theorem, for the Feynman gauge propagator, we arrive at\footnote{Note that, unlike for the fermion propagator, conjugation does not change the argument sign in $D_{\mu\nu}$, c.f. eq.~\eqref{eq:conjugate_fermion_propagator}.}
\begin{eqnarray}\label{eq:em_field_separation}
    \nonumber D_{\mu\nu}(x-y) = \theta(x^0-y^0)D_{\mu\nu}^{(+)}(x-y) + \theta(y^0-x^0)D_{\mu\nu}^{(-)}(x-y), \\
    D_{\mu\nu}^*(x-y) = \theta(x^0-y^0)D_{\mu\nu}^{(-)}(x-y) + \theta(y^0-x^0)D_{\mu\nu}^{(+)}(x-y),
\end{eqnarray}
where
\begin{align}\label{eq:Wightman_electromagnetic}
    D_{\mu\nu}^{(\pm)}(x-y) = -g_{\mu\nu}\int \frac{d^4 l}{(2\pi)^3}\theta(\pm l^0)\delta(l^2)e^{-il(x-y)}
    = \int \frac{d^3\bm{l}}{(2\pi)^3} \frac{-g_{\mu\nu}}{2\omega_{\bm{l}}}\left.e^{\mp il(x-y)}\right|_{l^0 = \omega_{\bm{l}} = |\bm{l}|}
\end{align}
are the Wightman functions for the electromagnetic field. There are some subtleties related to gauge invariance, to be discussed later.

Consider a Feynman diagram $G$ with $n$ vertices $\{x_i\}_{i=1}^n$, $n_e$ ingoing and $n_e$ outgoing fermion lines,\footnote{Here, terms ``ingoing'' and ``outgoing'' refer to the direction of the fermion lines, and not of the particles' momenta. For example, an ingoing positron is represented by an outgoing fermion line. Due to charge conservation, numbers of ingoing and outgoing lines defined this way, obviously, match.} as well as $\ngin$ ingoing photon lines and $\ngout$ outgoing photon lines. Following Veltman \cite{veltman1994diagrammatica}, we introduce an auxiliary function 
$$F_{ab}^{\mu\nu}(x_1,...,x_n;G),$$
with tensor and spinor multi-indices $\mu = (\mu_1...\mu_{\ngout})$, $\nu = (\nu_1...\nu_{\ngin})$, and $a = (a_1...a_{n_e})$, $b = (b_1...b_{n_e})$. Function $F^{\mu\nu}_{ab}$ represents the diagram $G$ according to standard Feynman rules, but with omitted bispinor amplitudes and polarization vectors corresponding to the in- and outgoing lines.

Next, we introduce the operation of circling a vertex by the additional set of rules:
\begin{itemize}
    \item An internal line connecting an uncircled vertex $x_k$ to a circled vertex $x_i$ corresponds to the positive-energy Wightman function 
    \begin{equation}
        \includegraphics[align=c,scale=2]{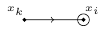} = S^{(+)}(x_i-x_k), \quad \includegraphics[align=c,scale=2]{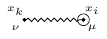} = D^{(+)}_{\mu\nu}(x_i-x_k);
    \end{equation}
    \item An internal line connecting a circled vertex $x_k$ to an uncircled vertex $x_i$ corresponds to the negative-energy Wightman function 
    \begin{equation}
        \includegraphics[align=c,scale=2]{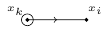} = S^{(-)}(x_i-x_k), \quad \includegraphics[align=c,scale=2]{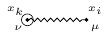} = D^{(-)}_{\mu\nu}(x_i-x_k);
    \end{equation}
    \item An internal line connecting a circled vertex $x_k$ to a circled vertex $x_i$ corresponds to the conjugated propagator
    \begin{equation}
        \includegraphics[align=c,scale=2]{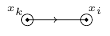} = \bar{S}(x_k-x_i), \quad \includegraphics[align=c,scale=2]{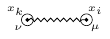} = D^{*}_{\mu\nu}(x_i-x_k);
    \end{equation}
    \item To each circled vertex, an additional factor $(-1)$ is assigned:
    \begin{equation}
        \includegraphics[align=c,scale=2]{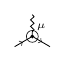} = +ie\gamma^\mu.
    \end{equation}
\end{itemize}
Assume we circle some vertices in diagram $G$. Then we associate the following function to such $G$:
$$F_{ab}^{\mu\nu}(x_{i_1},...,x_{i_m}|x_{i_{m+1}},...,x_{i_n}, G),$$ where the coordinates on the right side of the vertical line correspond to the circled vertices. 

Now, let us suppose that all time components $\{x_i^0\}_{i=1}^n$ have different values and let $x_k^0$ be the largest of them. Then, in virtue of relations \eqref{eq:spinor_field_free_separation} and \eqref{eq:em_field_separation}, circling vertex $x_k$ with the largest time results only in multiplying the corresponding function $F_{ab}^{\mu\nu}$ by the overall factor of $(-1)$. Furthermore, the set of all possible vertex circlings in $G$ splits into two equally sized subsets: with the largest time vertex circled and uncircled, respectively. For each diagram in the first subset, there is a diagram in the second subset that differs only by the circling of the largest time vertex; therefore, it is negative to the former. Hence, it follows that
\begin{equation}\label{eq:circling_equation0}
    \sum_{\text{circlings}}F_{ab}^{\mu\nu}(...|...;G) = 0,
\end{equation}
where the sum is taken over all possible vertex circlings. As an illustration, consider a loop diagram with two vertices $x_1$ and $x_2$. If $x_1^0 > x_2^0$:
\begin{equation*}
    \underbrace{\includegraphics[align=c,scale=1.5]{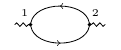} + \includegraphics[align=c,scale=1.5]{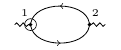}}_{=0} + \underbrace{\includegraphics[align=c,scale=1.5]{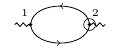} + \includegraphics[align=c,scale=1.5]{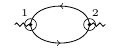}}_{=0} = 0
\end{equation*}       
If $x_1^0 < x_2^0$:
\begin{equation*}
    \underbrace{\includegraphics[align=c,scale=1.5]{circling_rules/pol_op.pdf} + \includegraphics[align=c,scale=1.5]{circling_rules/pol_op_circ_2.pdf}}_{=0} + \underbrace{\includegraphics[align=c,scale=1.5]{circling_rules/pol_op_circ_1.pdf} + \includegraphics[align=c,scale=1.5]{circling_rules/pol_op_circ_both.pdf}}_{=0} = 0
\end{equation*}

Note that to diagram $G$ with all vertices circled corresponds the function
\begin{equation}
    \begin{split}
        F_{ab}^{\mu\nu}(|x_1,...x_n;G) &= \bar{F}_{ab}^{\mu\nu}(x_1,...,x_n;\bar{G}) \\
        & \equiv \gamma^0_{a_1 c_1}...\gamma^0_{a_{n_{e}} c_{n_{e}}}\left(F^{\mu\nu}(x_1,...x_n;\bar{G})\right)^\dagger_{cd}\gamma^0_{d_1 b_1}...\gamma^0_{d_{n_{e}} b_{n_{e}}}
    \end{split}
\end{equation}
where diagram $\bar{G}$ differs from $G$ by that all lines are inverted. Therefore, we arrive at
\begin{equation}\label{eq:circling_equation1}
    F_{ab}^{\mu\nu}(x_1,...,x_n;G) + \bar{F}_{ab}^{\mu\nu}(x_1,...,x_n;\bar{G}) = -\sum_{\text{circlings}\backslash\{\emptyset,\text{all}\}}F_{ab}^{\mu\nu}(...|...;G).
\end{equation}
\begin{figure}
    \centering
    \begin{subfigure}{0.45\textwidth}
        \includegraphics[width=\textwidth]{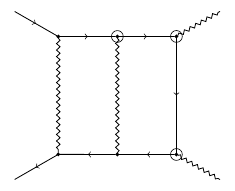}
        \caption{Circling the vertices.}
        \label{fig:cVc_circlings}
    \end{subfigure}
    \hfill
    \begin{subfigure}{0.45\textwidth}
        \includegraphics[width=\textwidth]{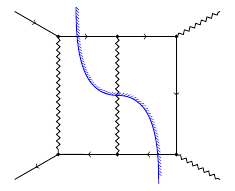}
        \caption{Cutting the diagram.}
        \label{fig:cVc_cuts}
    \end{subfigure}

    \begin{subfigure}{0.45\textwidth}
        \includegraphics[width=\textwidth]{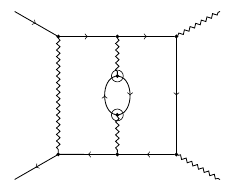}
        \caption{Circling the vertices.}
        \label{fig:cVc_circlings_isolated}
    \end{subfigure}
    \hfill
    \begin{subfigure}{0.45\textwidth}
        \includegraphics[width=\textwidth]{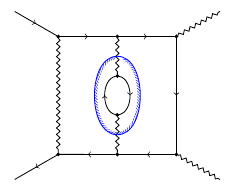}
        \caption{Cutting the diagram.}
        \label{fig:cVc_cuts_loop}
    \end{subfigure}
    \caption{An illustration of the equivalence between circling the vertices and cutting the diagram.}
    \label{fig:cVc}
\end{figure}
Circling the vertices basically divides the vertices into two different groups. The same can be achieved by cutting the diagram with one or more continuous sections that divide it into ``shadowed'' and ``unshadowed'' regions. Such a cut can cross each internal line only once and cannot cross any of the external lines. All vertices in the shadowed regions of the cut are circled, and all vertices in the unshadowed regions are uncircled. A cut can either be ``open``, i.e. its ends extend out of the diagram (see Figs.~\ref{fig:cVc_circlings} and \ref{fig:cVc_cuts}), or form a closed loop inside the diagram (see Figs.~\ref{fig:cVc_circlings_isolated} and \ref{fig:cVc_cuts_loop}). Since any two open cuts can be connected outside the diagram, there may only be one open cut and an arbitrary number of closed loops (bounded only by the number of vertices). In what follows, we will stick to cutting instead of circling. We can then rewrite the previous equation as
\begin{equation}\label{eq:cutting_equation}
    F_{ab}^{\mu\nu}(x_1,...,x_n;G) + \bar{F}_{ab}^{\mu\nu}(x_1,...,x_n;\bar{G}) = -\sum_{\text{cuttings}}F_{ab}^{\mu\nu}(...|...;G),
\end{equation}
where the sum is now taken over all possible ways to cut the internal lines in $G$. Expression \eqref{eq:cutting_equation} is referred to as the \textit{cutting equation} \cite{veltman1994diagrammatica}. The set of prescriptions that define the operation of circling vertices or, equivalently, cutting the propagators is known as the \textit{cutting rules}.

\section{Cutting rules in QED in a plane-wave background}
\label{sec:iii}
\subsection{Furry picture framework}\label{sec:iii_1}
In the presence of an external classical (background) field $\mathcal{A}(x)$, the QED Lagrangian changes to \cite{Di_Piazza_2012,fedotov2023high}
\begin{equation}
    \mathcal{L} = -\frac{1}{4}F_{\mu\nu}F^{\mu\nu} - \frac{1}{2\xi}(\partial_\mu A^\mu)^2 + \bar{\psi}(i\slashed{\partial} -e\slashed{\mathcal{A}} - m)\psi - e\bar{\psi}\slashed{A}\psi.
\end{equation}
If the external field does not produce $e^-e^+$ pairs spontaneously from the vacuum, the one-particle fermion states are well-defined and can be found as a complete set of positive-energy solutions to the Dirac equation in the external field
\begin{equation}\label{eq:Dirac_equation_with_field}
    (i\slashed{\partial} - e\slashed{\mathcal{A}} - m)\psi = 0.
\end{equation}
In what follows, we consider a plane-wave background
\begin{equation}
    \mathcal{A} = \mathcal{A}(\varphi),
\end{equation}
with phase $\varphi = kx$ and wave vector $k$ satisfying
\begin{equation}
    k^2 =0, \quad k \mathcal{A}(\varphi) = 0.
\end{equation}
In this case, the positive and negative energy solutions to eq.~\eqref{eq:Dirac_equation_with_field} are given by the Volkov functions \cite{volkov1935class}
\begin{equation}
    \label{eq:psi_furry}
    \psi_{p,\sigma}^{(+)}(x) = E_p(x)u(\bm{p},\sigma), \quad \psi_{p,\sigma}^{(-)}(x) = E_{-p}(x)v(\bm{p},\sigma),
\end{equation}
where
\begin{equation}\label{eq:Ep_function_general}
    E_p(x) = \left(1+\frac{e\slashed{k} \slashed{\mathcal{A}}(kx)}{2 \, kp}\right) \exp\left[ -ipx -ie\int_0^{kx}\left(\frac{p\mathcal{A}(\varphi)}{kp}-\frac{e\mathcal{A}^2(\varphi)}{2\, kp}\right)d\varphi \right]
\end{equation}
and $u(\bm{p},\sigma)$ and $v(\bm{p},\sigma)$ satisfy eqs.~\eqref{eq:u_v_properties}. Modes $E_p(x)$ form a complete set of orthogonal functions \cite{ritus1985quantum}:
\begin{align}
    &\int d^4x E_p(x)\bar{E}_{p'}(x) = (2\pi)^4\delta(p'-p), \\
    &\int\frac{d^4p}{(2\pi)^4}E_p(x)\bar{E}_p(y) = \delta(x-y).
\end{align}
Even though the 4-vector $p$ now differs from the kinetic particle momentum (which is not conserved), we refer to it as ``momentum'' for brevity. $E_p$-functions serve as Fourier modes, and we can use them to map the operators defined in the coordinate space to the $p$-space. The free fermion field operators in the Furry picture\footnote{Those are dressed exactly by the external field but without interaction with the quantized electromagnetic field.} 
are given by
\begin{align}
    &\psi(x) = \int \frac{d^3\bm{p}}{(2\pi)^3} \frac{1}{\sqrt{2\varepsilon_{\bm{p}}}} \sum_{\sigma=1}^2 \left(a_{\bm{p},\sigma} E_p(x) u(\bm{p}, \sigma) + b_{\bm{p},\sigma}^\dagger E_{-p}(x) v(\bm{p}, \sigma)\right), \\
    &\bar{\psi}(x) = \int \frac{d^3\bm{p}}{(2\pi)^3} \frac{1}{\sqrt{2\varepsilon_{\bm{p}}}} \sum_{\sigma=1}^2 \left(b_{\bm{p},\sigma} \bar{v}(\bm{p}, \sigma)\bar{E}_{-p}(x) + a_{\bm{p},\sigma}^\dagger \bar{u}(\bm{p}, \sigma) \bar{E}_p(x)\right),
\end{align}
where $p^0 = \varepsilon_{\bm{p}} = \sqrt{\bm{p}^2+m^2}$. The external field-dressed Feynman propagator for the spin-$\frac{1}{2}$ field is given by
\begin{equation}\label{eq:propagator_definition_Ep}
    S(x,y) = \int \frac{d^4 p}{(2\pi)^4} E_p(x) \frac{i(\slashed{p}+m)}{p^2-m^2+i0} \bar{E}_p(y).
\end{equation}
The key difference from the ordinary QED is that the fermion propagator no longer depends solely on $x-y$. In diagrams, the field-dressed propagator is represented by a double line, which is equivalent to an all-order resummation of the interactions with the external field:
\[\includegraphics[align=c,height=1cm]{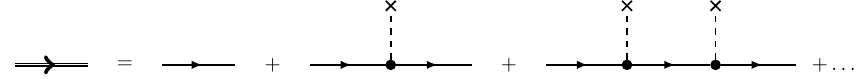}.\]
As a photon is neutral, its propagator is still given by eq.~\eqref{eq:em_field_feynman_propagator}. 

Instead of using coordinate-space Feynman rules, we can pass to the  $E_p$-representation, where the propagators are diagonal and take the same form as in the momentum representation in ordinary QED:
\begin{equation}
    S(p) = \frac{i}{p^2-m^2+i\varepsilon}, \quad D_{\mu\nu}(l) = \frac{-ig_{\mu\nu}}{l^2+i\varepsilon}.
\end{equation}
However, this comes at a cost of dressing the vertices due to the interaction with the background \cite{ritus1972vacuum,Di_Piazza_2012,popruzhenko2023dynamics}:
\begin{align}
    \int d^4 x \bar{E}_{p'}(x) \gamma^\mu E_p(x)e^{-ilx} = \int_{-\infty}^{+\infty} ds \, \Gamma^{\mu}(s;p', p)\,(2\pi)^4\delta^{(4)}(p'-p-l-sk),
\end{align}
where 
\begin{align}\label{eq:dressed_vertex_general}
    \Gamma^{\mu}(s; p', p) =& \int_{-\infty}^{+\infty} \frac{d\varphi}{2\pi} \left[1+\frac{e \slashed{\mathcal{A}}(\varphi) \slashed{k} }{2 \, kp'}\right] \gamma^{\mu} \left[1+\frac{e\slashed{k} \slashed{\mathcal{A}}(\varphi)}{2 \, kp}\right] \nonumber \\ & \times \exp\left[ is\varphi + ie\int_0^{\varphi} \left(\frac{p'\mathcal{A}(\phi)}{kp'}-\frac{p\mathcal{A}(\phi)}{kp}-\frac{e\mathcal{A}^2(\phi)}{2\, kp'}+\frac{e\mathcal{A}^2(\phi)}{2\, kp}\right)d\phi \right],
\end{align}
is the induced nonlocal vertex in the momentum representation.

Momentum-space Feynman rules are formulated as follows. Consider a diagram $G$ for a scattering process with $V$ vertices. Assign a real number $s_j$ and a factor $-ie\Gamma^{\mu_j}(s_j;p'_j,p_j)$ to each vertex, and apply momentum conservation in the form
\begin{equation}\label{eq:momentum_conservation_sfqed}
    p'_j = p_j+l_j+s_j k,
\end{equation}
where $p_j$ and $p'_j$ are the fermion particle momenta going into and out of the $j$-th vertex,  $l_j$ is the ingoing photon momentum and the last term accounts for the momentum gained from the external field at the vertex. The remaining rules are the same as in ordinary QED.  The contribution of the diagram $G$ to the $\mT$-matrix\footnote{The nontrivial part of the $S$-matrix: $S =1 + i\mT$.} element is given by
\begin{equation}\label{eq:s_matrix_sfqed}
    \mT_{fi}(G) = \int d s_1 ...ds_V \, \mathcal{M}(s_1,...,s_V) \, (2\pi)^4\delta\left(P_i + k\sum_{j=1}^V s_j - P_f\right),
\end{equation}
where $\mathcal{M}(s_1, \dots, s_V)$ abbreviates the partial amplitude, $\ket{i}$ and $\ket{f}$ denote the initial and final states, $P_i$ and $P_f$ denote the total incoming and outgoing momenta. While taking the mod-square of the matrix element, we  encounter a product of delta functions of the form \cite{ritus1985quantum}
\begin{align}\label{eq:delta_funcs_product}
    \delta(P_i + sk- P_f) \, \delta(P_i + t k - P_f) &= \delta\left((s-t)k\right) \, \delta(P_i + s k - P_f) \nonumber \\ 
    & = \left.\frac{\delta((s-t)k)}{\delta(s-t)}\right|_{s=t}\,\delta(s-t)\,\delta(P_i+sk-P_f) \nonumber \\
    & = \frac{V T}{(2\pi)^3L_{\varphi}}\,\delta(s-t)\delta(P_i+sk-P_f),
\end{align}
where $VT$ is the large space-time volume, $L_\varphi$ is the corresponding large interval of the phase $\varphi = kx$ and we have used
\begin{align}\label{eq:L_phi_def}
    &\left.\delta(s-t)\right|_{s=t} = \left.\int_{-\infty}^{+\infty}\frac{d\varphi}{2\pi} e^{i(s-t)\varphi}\right|_{s=t} \simeq \left.\int_{-L_\varphi/2}^{L_\varphi/2}\frac{d\varphi}{2\pi} e^{i(s-t)\varphi}\right|_{s=t} = \frac{L_{\varphi}}{2\pi}.
\end{align}

In a reference frame such that $k^{\mu} = (\omega, 0, 0, \omega)$, we have
\begin{equation}
    kp =  \omega p_-,
\end{equation}
where $p_-=p^0-p^3$ is the light-front momentum component. We will occasionally also use another notation: $p_+ = (p^0+p^3)/2$. Since $k^2 = 0$, from eq.~\eqref{eq:momentum_conservation_sfqed} it follows that the net light-front momentum is conserved at the vertex,
\begin{equation}
    p'_{j,-} = p_{j,-} + l_{j,-}.
\end{equation}

In sections~\ref{sec:v},~\ref{sec:vi} and~\ref{sec:vii}, we perform calculations in the special case of a CCF. In this configuration, the background field can be conveniently expressed as
\begin{equation}
    \mathcal{A}_\mu(\varphi) = a_\mu \varphi,
\end{equation}
where $a_\mu$ is a constant vector, satisfying $ka = 0$. Hence, we can integrate over $\phi$ in eq.~\eqref{eq:dressed_vertex_general}, so that the dressed vertices are given by
\begin{equation}\label{eq:Gamma_cn}
    \Gamma^{\mu}(s;p',p) = \gamma^{\mu}C_0(s; p', p) + \left( \frac{e \slashed{a} \slashed{k}}{2\,kp'}\gamma^{\mu} + \gamma^{\mu} \frac{e \slashed{k} \slashed{a}}{2\, kp} \right) C_1(s;p',p) -\frac{e^2 a^2 k^{\mu} \slashed{k}}{2 (kp')(kp)}\,C_2(s;p',p),
\end{equation}
where the coefficients (formfactors) $C_n(s;p',p)$ are expressed in terms of Airy functions in appendix~\ref{sec:app_c}.

\subsection{The cutting equation}
\label{sec:iii_2}
Consider the Feynman propagator for the spin-$\frac{1}{2}$ field in the Furry picture. By definition
\begin{equation}\label{eq:propagator_definition_furry}
    \begin{split}
    S_{ab}(x,y) \equiv & \langle 0|T\psi_a(x)\bar{\psi}_b(y)|0\rangle \\ = & \,\theta(x^0-y^0)\bra{0}\psi_a(x)\bar{\psi}_b(y)\ket{0} - \theta(y^0-x^0)\bra{0}\bar{\psi}_b(y)\psi_a(x)\ket{0} \\
    \equiv & \,\theta(x^0-y^0)S_{ab}^{(+)}(x,y) + \theta(y^0-x^0)S_{ab}^{(-)}(x,y),
    \end{split}
\end{equation}
where the Wightman functions are given by
\begin{gather}
    \begin{split}
    S^{(+)}(x,y) & = \int \frac{d^3 \bm{p}}{(2\pi)^3} \frac{1}{2\varepsilon_{\bm{p}}}\left.\sum_\sigma \psi_{p,\sigma}^{(+)}(x)\bar{\psi}_{p,\sigma}^{(+)}(y)\right|_{p^0=\varepsilon_{\bm{p}}} \\
    & = \int \frac{d^4 p}{(2\pi)^3}\theta(p^0)\delta(p^2-m^2)E_p(x)(\slashed{p}+m)\bar{E}_p(y), 
    \end{split}
    \label{eq:Wightman_spinor_field_sfqed_plus}
    \\
    \begin{split}
    S^{(-)}(x,y) & = -\int \frac{d^3 \bm{p}}{(2\pi)^3} \frac{1}{2\varepsilon_{\bm{p}}}\left.\sum_\sigma \psi_{p,\sigma}^{(-)}(x)\bar{\psi}_{p,\sigma}^{(-)}(y)\right|_{p^0=\varepsilon_{\bm{p}}} \\
    & = \int \frac{d^4 p}{(2\pi)^3}\theta(-p^0)\delta(p^2-m^2)E_p(x)(\slashed{p}+m)\bar{E}_p(y).
    \end{split}
    \label{eq:Wightman_spinor_field_sfqed_minus}
\end{gather}
and $\psi_{p,\sigma}$, $\bar{\psi}_{p,\sigma}$ are now given by eq.~\eqref{eq:psi_furry}.\footnote{Let us note here that the representations of the Volkov propagator given in eqs.~\eqref{eq:propagator_definition_Ep} and \eqref{eq:propagator_definition_furry} are generally considered equivalent \cite{ritus1985quantum, brown1964interaction, di2018completeness}. However, we are not aware of an explicit proof of this fact in the literature that would connect the two expressions by a contour integration in the $p^0$-plane with a consistent treatment of the singularity at $p_-=0$. When formulating the cutting rules, we rely only on eq.~\eqref{eq:propagator_definition_furry}, which is the fundamental definition of the electron propagator underlying the perturbation theory formulation in the canonical equal-time quantization \cite{ritus1985quantum, fradkin1991book}.}

By using the properties of $E_p$-functions, it is straightforward to show that Wightman functions still satisfy
\begin{equation}
    \bar{S}^{(\pm)}(y,x)  = S^{(\pm)}(x,y).
\end{equation}
Therefore, inherently from ordinary QED, we still have the key relations underlying the cutting equation derivation:
\begin{eqnarray}\label{eq:spinor_field_sfqed_separation}
    \nonumber S(x,y) = \theta(x^0-y^0)S^{(+)}(x,y) + \theta(y^0-x^0)S^{(-)}(x,y), \\
    \bar{S}(y,x) = \theta(x^0-y^0)S^{(-)}(x,y) + \theta(y^0-x^0)S^{(+)}(x,y).
\end{eqnarray}
As a result, the cutting equation given in eq.~\eqref{eq:cutting_equation} holds in a QED in a plane-wave background with the following cutting rules:
\begin{itemize}
    \item To an internal line in the unshadowed region corresponds the Feynman propagator:
    \begin{equation}
        \includegraphics[align=c,scale=2]{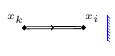} = S(x_i,x_k), \quad \includegraphics[align=c,scale=2]{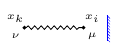} = D_{\mu\nu}(x_i-x_k);
    \end{equation}
    \item To an internal line going from the unshadowed to the shadowed region corresponds the positive-energy Wightman function:
    \begin{equation}
        \includegraphics[align=c,scale=2]{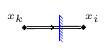} = S^{(+)}(x_i,x_k), \quad \includegraphics[align=c,scale=2]{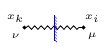} = D_{\mu\nu}^{(+)}(x_i-x_k);
    \end{equation}
    \item To an internal line going from the shadowed to the unshadowed region corresponds the negative-energy Wightman function:
    \begin{equation}
        \includegraphics[align=c,scale=2]{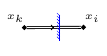} = S^{(-)}(x_i,x_k), \quad \includegraphics[align=c,scale=2]{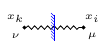} = D_{\mu\nu}^{(-)}(x_i-x_k);
    \end{equation}
    \item To an internal line in the shadowed region corresponds the conjugated propagator:
    \begin{equation}
        \includegraphics[align=c,scale=2]{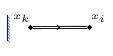} = \bar{S}(x_k,x_i), \quad \includegraphics[align=c,scale=2]{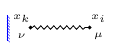} = D^*_{\mu\nu}(x_i-x_k);
    \end{equation}
    \item To each vertex in the shadowed region is assigned an additional factor  $(-1)$: 
    \begin{equation}
        \includegraphics[align=c,scale=2]{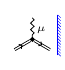} = -ie\gamma^\mu, \quad \includegraphics[align=c,scale=2]{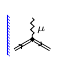} = +ie\gamma^\mu.
    \end{equation}
\end{itemize}

The next step is expressing the cutting equation in the form of a relation between $\mT$-matrix elements. Following Feynman rules for Furry picture QED, if an in- or outgoing line is attached to vertex $x_i$, we multiply  eq.~\eqref{eq:cutting_equation}:
\begin{itemize}
    \item For an ingoing fermion particle with 4-momentum $p$ and spin state $\sigma$ --- by \newline $(E_p(x_i)u(\bm{p},\sigma))_{b_i}$;
    \item For an outgoing fermion particle with 4-momentum $p$ and spin state $\sigma$ --- by \newline $(\bar{u}(\bm{p},\sigma)\bar{E}_p(x_i))_{a_i}$;
    \item For an ingoing fermion anti-particle with 4-momentum $p$ and spin state $\sigma$ --- by $(\bar{v}(\bm{p},\sigma)\bar{E}_{-p}(x_i))_{a_i}$;
    \item For an outgoing fermion anti-particle with 4-momentum $p$ and spin state $\sigma$ --- by $(E_{-p}(x_i)v(\bm{p},\sigma))_{b_i}$;
    \item For an ingoing photon with momentum $l$ and spin state $\lambda$ --- by $\epsilon_{\nu_i}(\bm{l},\lambda)e^{-ilx_i}$;
    \item For an outgoing photon with momentum $l$ and spin state $\lambda$ --- by $\epsilon^*_{\nu_i}(\bm{l},\lambda)e^{ilx_i}$.
\end{itemize}
After that, we contract all spinor and tensor indices and integrate over all $x_i$. The expression should be multiplied by a fermion statistical factor. Each closed fermion loop contributes a factor of ($-1$), while an additional ($\pm 1$) factor might come from interchanging the external lines \cite{schwartz2014quantum}. Finally, we have to divide the result by the symmetry factor $g_G$ of the diagram. The LHS of the cutting equation is turned into
\begin{equation}\label{eq:cutting_equation_amplitudes_LHS}
   i\mT_{fi}(G) - i\mT_{if}^*(\bar{G}), 
\end{equation}
where $\mT_{if}(\bar{G})$ is the contribution of the inverted diagram\footnote{A diagram with all internal and external lines inverted.} $\bar{G}$ to the $\mT$-matrix element for the $\ket{f} \rightarrow \ket{i}$ process.

Notice that, in the above derivation of the cutting equation, we have assumed that all vertices have different times. However, function $F_{ab}^{\mu\nu}(x;G)$ is nonsingular at matching time points. Therefore, such points do not contribute to the $\mT$-matrix, and the resulting cutting equation holds after integration anyway. 

Let us apply the same steps to the RHS of the cutting equation. A cut splits the whole diagram $G$ into two or more connected subdiagrams, each lying on either the shadowed or unshadowed side of the cut. Note that for on-shell momenta
\begin{equation}
    \theta(p_-) = \theta(p^0).
\end{equation}
Therefore, according to eqs.~\eqref{eq:Wightman_electromagnetic}, \eqref{eq:Wightman_spinor_field_sfqed_plus}, \eqref{eq:Wightman_spinor_field_sfqed_minus}, a cut requires positive flow of the light-front momentum from the unshadowed to the shadowed side of the cut. Recall that the net light-front momentum is conserved in a plane-wave field. It follows that the cut diagrams with isolated regions give zero contribution. By an isolated region we mean a region of a diagram, that can not be connected to any of the external lines without crossing the cut. By the same token, cut diagrams containing a subdiagram on the unshadowed side of the cut connected only to outgoing lines (or vice versa) give zero contribution as well. In other words, for a cut diagram to give a nontrivial contribution to the RHS of the cutting equation, each connected subdiagram on the unshadowed side of the cut must be connected to at least one ingoing line and each connected subdiagram on the shadowed side of the cut must be connected to at least one outgoing line. 

\begin{figure}
    \centering
    \begin{subfigure}{0.45\textwidth}
            \includegraphics[width=\textwidth]{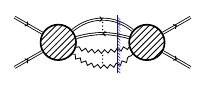}
        \caption{}
        \label{fig:2to2_2cut_a}
    \end{subfigure}
    \hfill
    \begin{subfigure}{0.45\textwidth}
            \includegraphics[width=\textwidth]{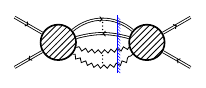}
        \caption{}
        \label{fig:2to2_2cut_b}
    \end{subfigure}

    \begin{subfigure}{0.45\textwidth}
        \includegraphics[width=\textwidth]{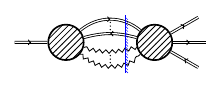}
        \caption{}
        \label{fig:2to2_2cut_c}
    \end{subfigure}
    \hfill
    \begin{subfigure}{0.45\textwidth}
        \includegraphics[width=\textwidth]{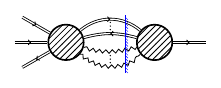}
        \caption{}
        \label{fig:2to2_2cut_d}
    \end{subfigure}
    \caption{Different nontrivial cuts of the $e^-e^-\rightarrow e^-e^-$ diagram that split it into two connected subdiagrams.}
    \label{fig:2to2_2cuts}
\end{figure}

\begin{figure}
    \centering
    \begin{subfigure}{0.45\textwidth}
        \includegraphics[width=\textwidth]{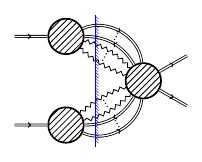}
        \caption{}
        \label{fig:2to2_3cut_a}
    \end{subfigure}
    \hfill
    \begin{subfigure}{0.45\textwidth}
        \includegraphics[width=\textwidth]{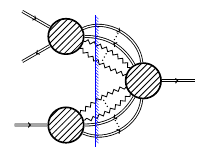}
        \caption{}
        \label{fig:2to2_3cut_b}
    \end{subfigure}

    \begin{subfigure}{0.45\textwidth}
        \includegraphics[width=\textwidth]{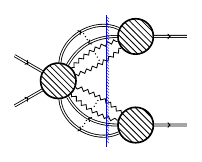}
        \caption{}
        \label{fig:2to2_3cut_c}
    \end{subfigure}
    \hfill
    \begin{subfigure}{0.45\textwidth}
        \includegraphics[width=\textwidth]{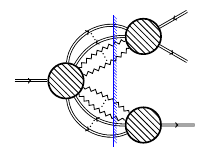}
        \caption{}
        \label{fig:2to2_3cut_d}
    \end{subfigure}
    \caption{Different nontrivial cuts of the $e^-e^-\rightarrow e^-e^-$ diagram that split it into three connected subdiagrams.}
    \label{fig:2to2_3cuts}
\end{figure}

\begin{figure}
    \centering
    \includegraphics[width=0.45\textwidth]{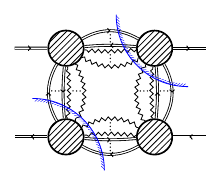}
    \caption{Nontrivial cut of the $e^-e^-\rightarrow e^-e^-$ diagram that splits it into four connected subdiagrams.}
    \label{fig:2to2_4cut}
\end{figure}

\begin{figure}  
    \centering
    \includegraphics[width=0.45\textwidth]{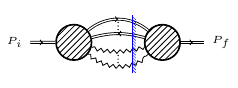}
    \caption{Nontrivial cut of the diagram for the $1\rightarrow1$ process.}
    \label{fig:1to1_cut}
\end{figure}

As an illustration, consider a general diagram for the $e^-e^- \rightarrow e^-e^-$ process in a plane-wave background. Nontrivial cuts are the single continuous open cuts that divide it into two (figure~\ref{fig:2to2_2cuts}), three (figure~\ref{fig:2to2_3cuts}), and four (figure~\ref{fig:2to2_4cut}) connected subdiagrams. If a cut divided the diagram into more than four subdiagrams, at least one of them would be isolated. Therefore, such cuts do not contribute to the cutting equation. Note that ingoing and outgoing lines can appear both on the shadowed and unshadowed sides of the cut. In the current paper we avoid these complications and consider only single-particle initial and final states. In that case nontrivial cuts divide the diagram into two subdiagrams, as illustrated in figure~\ref{fig:1to1_cut}. We denote these subdiagrams $A$ and $B$, the former on the unshadowed side of the cut and the latter on the shadowed side.

According to definitions \eqref{eq:Wightman_electromagnetic}, \eqref{eq:Wightman_spinor_field_sfqed_plus}, \eqref{eq:Wightman_spinor_field_sfqed_minus}, when an internal line is cut, the corresponding virtual particle goes on shell. The photon line direction can always be chosen such that all virtual photons going on shell have positive energy. The spinor field case is different: As it follows from eqs.~\eqref{eq:Wightman_spinor_field_sfqed_plus}, \eqref{eq:Wightman_spinor_field_sfqed_minus}, when a spinor propagator is replaced with the negative-energy Wightman function, it corresponds to an on-shell anti-particle with positive energy and yields an additional factor of $(-1)$.

Each of the subdiagrams corresponds to the $\mT$-matrix element multiplied by the symmetry  and statistical fermion factors of the subdiagram. The subdiagram $B$ on the shadowed side actually corresponds to the complex-conjugate $\mT$-matrix element for the inverted subdiagram $\bar{B}$. Integrals over spatial components of the momenta form the integral over the Lorentz-invariant phase volume of an intermediate state $\ket{X}$ of virtual particles put on shell, and the result is summed over spin states of the intermediate particles. Finally, statistical factors, associated with initial and final states $\ket{i}$ and $\ket{f}$, are identical in LHS and RHS of the cutting equation and cancel out. Statistical factor, associated with the state $\ket{X}$, is squared and does not contribute. This way we arrive at
\begin{equation}\label{eq:cutting_equation_amplitudes_sfqed}
    \begin{split}
        i\mT_{if}^*(\bar{G}) - i\mT_{fi}(G)  = \sum_{\text{cuttings}} \frac{g_A \cdot g_B \cdot N_X}{g_G} & (-1)^{L(G)-L(A)-L(B)+n_{\text{a.p.}}(X)} \\
        &\sum_{\text{spin}} \int  d\Pi_X \, \mT_{Xi}(A) \, \mT_{Xf}^*(\bar{B}),
    \end{split}
\end{equation}
where
\begin{equation}
    \label{eq:measure_Pi_tilde}
    d\Pi_X =  \frac{1}{N_X} \prod_{k=1}^{n_e(X)} \frac{d^3 \bm{p}_k}{(2\pi)^3 2\varepsilon_{\bm{p}_k}} \prod_{k=1}^{n_\gamma(X)} \frac{d^3 \bm{l}_k}{(2\pi)^3 2\omega_{\bm{l}_k}}.
\end{equation}
 is the invariant phase volume of the intermediate state $\ket{X}$ with  $n_e(X)$ fermions and $n_\gamma(X)$ photons, $N_X$ is the number of rearrangements of identical intermediate particles, $n_{\text{a.p.}}(X)$ is the number of intermediate anti-particles, $L(G), L(A)$ and $L(B)$ are the numbers of fermion loops in diagram $G$ and its subdiagrams $A$ and $B$, respectively. Finally, $g_G$, $g_A$, and $g_B$ are the symmetry factors of diagram $G$ and its subdiagrams $A$ and $B$, respectively.\footnote{In spinor QED, however, any connected diagram for a scattering process has a trivial symmetry factor $g=1$ \cite{hue2012general}.}

Two major comments are in order. First, in our derivation of the RHS of  eq.~\eqref{eq:cutting_equation_amplitudes_sfqed}, all intermediate particles go to a bare mass shell, which is fine in the lowest order of perturbation theory, as is of our primary interest here. However, when higher-order self-energy corrections to the cut lines on both sides of the cut are properly resummed, proper renormalization constants show up, and the masses of intermediate particles shift to their physical values. Second, when deriving the RHS, we used eq.~\eqref{eq:WTidentity}, which is justified by the Ward-Takahashi identity. The latter, however, holds perturbatively only when applied to a gauge-invariant sum of diagrams of the given order. Therefore eq.~\eqref{eq:cutting_equation_amplitudes_sfqed} holds for either a gauge-invariant diagram or a sum of all gauge-dependent ones of the given order.

Finally, let us formulate the momentum-space (namely, $E_p$-space) version of the cutting rules. According to definitions \eqref{eq:Wightman_electromagnetic} \eqref{eq:Wightman_spinor_field_sfqed_plus}, \eqref{eq:Wightman_spinor_field_sfqed_minus}, 
the substitutions
$$S(x,y) \mapsto S^{(\pm)}(x,y) \quad \text{and} \quad D_{\mu\nu}(x-y) \mapsto D_{\mu\nu}^{(\pm)}(x-y)$$
are equivalent to
\begin{equation}\label{eq:cutting_rules_momentum_1}
    \frac{1}{p^2-m^2+i\varepsilon} \mapsto -2\pi i \,\theta(\pm p_-)\,\delta(p^2-m^2) \quad \text{and} \quad \frac{1}{l^2+i\varepsilon} \mapsto -2 \pi i \, \theta(\pm l_-) \, \delta(l^2),
\end{equation}
where we have used that $\theta(p_-) = \theta(p^0)$ on shell. The substitutions 
$$S(x,y) \mapsto \bar{S}(y,x) \quad \text{and} \quad D_{\mu\nu}(x-y) \mapsto D_{\mu\nu}^*(x-y)$$
are equivalent to complex conjugation:
\begin{equation}\label{eq:cutting_rules_momentum_2}
    \frac{i}{p^2-m^2+i\varepsilon} \mapsto \frac{-i}{p^2-m^2-i\varepsilon} \quad \text{and} \quad \frac{i}{l^2+i\varepsilon} \mapsto \frac{-i}{l^2-i\varepsilon}.
\end{equation}

\section{Illustration: second-order electron forward scattering}
\label{sec:iv}
As an illustration of the derived cutting equation and a prologue to the second topic of our paper, consider an electron forward scattering in a strong plane-wave background. At the leading order, the nontrivial part of the scattering matrix is given by the diagram:
\[\includegraphics[align=c,scale=1.5]{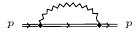}.\]

This diagram corresponds to a gauge-invariant $\mT$-matrix element, therefore we can apply equation \eqref{eq:cutting_equation_amplitudes_sfqed}, which takes the form\footnote{In equations,  diagrams represent the corresponding contributions to the $\mT$-matrix element.}
\begin{equation}\label{eq:1loop_cutting_eq}
    2\operatorname{Im}\left(\includegraphics[align=c,scale=1]{1loop_massop/1loop_massop_lab.pdf}\right) = \sum_{\text{spin}} \int d\Pi(q,l) \left|\includegraphics[align=c,scale=1]{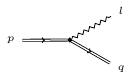}\right|^2
\end{equation}
and has already been proven by an explicit calculation of the LHS and RHS terms in this equation \cite{ritus1972radiative,ritus1985quantum}. Hereinafter, the summation over spin in the RHS of the cutting equation is applied to the final particle states. Averaging over the initial particle spin can be done, if needed, at the end of the calculation, however, we do not imply it by definition. 

At the next to leading order, we have three diagrams
\begin{equation}
    \label{eq:T4_diagrams}
    \mT^{(4)}_{pp} = \includegraphics[align=c,scale=1.3]{massop_bubble/2loop_massop_a_lab}+\includegraphics[align=c, scale=1.3]{massop_rainbow/2loop_massop_b_lab}+\includegraphics[align=c, scale=1.3]{massop_interference/2loop_massop_c_lab}. 
\end{equation}
First, let us focus on the first of them, which we denote $\mT_{pp}^{(4),\mathrm{pol}}$. This term will be relevant to our discussion in the follow-up sections. Let us emphasise that this contribution is gauge-invariant on shell by virtue of the transversality of the polarization operator (bubble) inserted into the photon line. Thus, we can apply the cutting equation. There are three nontrivial ways to cut such a diagram. One of them turns the diagram into a contribution to the so-called trident process:
\begin{equation}
 \label{eq:cut_trident}
 \begin{split}
 \includegraphics[align=c,scale=1]{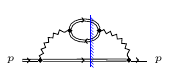} & = 2 \sum_{\text{spin}} \int d \Pi(q,g,f) \left|\includegraphics[align=c,scale=1]{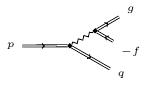}\right|^2 \\
 &= \sum_{\text{spin}} \int d \Pi(q,g,f) \left(\underset{\text{``direct'' diagram}\,\,\,}{\left|\includegraphics[align=c, scale=1]{trident/trident_dir_lab.pdf}\right|^2} + \underset{\text{``exchange'' diagram}\,\,\,}{\left|\includegraphics[align=c, scale=1]{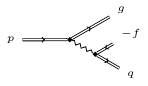}\right|^2} \right),
 \end{split}
\end{equation}
where we have used that ``direct'' and ``exchange'' diagrams equally contribute to the integral over the phase volume. Note that the factor $N_X=2$ comes from the number of rearrangements of two identical electrons in the cut. Cutting the loop yields the factor of $(-1)$, which is, however canceled by putting $n_{\text{a.p.}}(X)=1$ antiparticles (single positron) on shell. Symmetry factors equal to unity for both the loop and the resulting tree-level diagrams. Let us also demonstrate that eq.~\eqref{eq:cut_trident} holds by explicit application of the cutting rules at the level of matrix element formulation:
\begin{equation}
    \label{eq:cutting_equation_two_loop_explicit_amplitude}
    \begin{split}
        & \includegraphics[align=c,scale=1]{massop_bubble/2loop_massop_a_trident_cut_lab.pdf} =  \int dx'' dx' dy'' dy' \, \bar{\psi}_{p,\sigma}^{(+)}(x'')(+ie\gamma^\mu)S^{(+)}(x'',x')(-ie\gamma^\nu) \psi_{p,\sigma}^{(+)}(x') \\
        & \quad \times (-1)\operatorname{tr}\left[(+ie\gamma^\rho)S^{(+)}(y'',y')(-ie\gamma^\lambda)S^{(-)}(y',y'')\right] D^*_{\rho\mu}(y''-x'') D_{\lambda\nu}(y'-x') \\
        & = \int \frac{d^3 \bm{q}}{(2\pi)^3} \frac{1}{2\varepsilon_{\bm{q}}} \frac{d^3 \bm{g}}{(2\pi)^3} \frac{1}{2\varepsilon_{\bm{g}}} \frac{d^3 \bm{f}}{(2\pi)^3} \frac{1}{2\varepsilon_{\bm{f}}} \\
        & \quad \times \sum_{\sigma_1, \sigma_2, \sigma_3}  \int dx''dx'dy''dy' \, \bar{\psi}_{p,\sigma}^{(+)}(x'')(+ie\gamma^\mu)\psi_{q,\sigma_1}^{(+)}(x'') \, \bar{\psi}_{q,\sigma_1}^{(+)}(x')(-ie\gamma^\nu) \psi_{p,\sigma}^{(+)}(x') \\
        & \quad \times \operatorname{tr}\left[ (+ie\gamma^\rho)\,\psi^{(+)}_{g,\sigma_3}(y'') \bar{\psi}^{(+)}_{g,\sigma_3}(y')\,(-ie\gamma^\lambda)\,\psi^{(-)}_{f,\sigma_2}(y') \bar{\psi}^{(-)}_{f,\sigma_2}(y'') \right] D^*_{\rho\mu}(y''-x'') D_{\lambda\nu}(y'-x') \\
        & = 2 \sum_{\sigma_1, \sigma_2, \sigma_3} \int d\Pi(q,g,f) \\
        & \quad \times \left|\int dx' dy' \, \bar{\psi}_{q,\sigma_1}^{(+)}(x')(-ie\gamma^\nu) \psi_{p,\sigma}^{(+)}(x') \, \bar{\psi}^{(+)}_{g,\sigma_3}(y')(-ie\gamma^\lambda)\psi^{(-)}_{f,\sigma_2}(y') \, D_{\lambda\nu}(y'-x')\right|^2 \\
        &= 2 \sum_{\text{spin}} \int d \Pi(q,g,f) \left|\includegraphics[align=c,scale=1]{trident/trident_dir_lab.pdf}\right|^2,
    \end{split}
\end{equation}
where we have used that, according to definition \eqref{eq:measure_Pi_tilde}, the measure $d\Pi$ contains a factor of $1/N_X$ with $N_X=2$.

Two other ways to cut this diagram give a radiation correction to the nonlinear Compton effect:
\begin{equation}
\label{eq:cut_rad_corr}
\begin{split}
\includegraphics[align=c, scale=1]{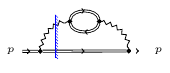} &+ \includegraphics[align=c, scale=1]{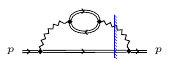} \\
&= 2 \operatorname{Re} \sum_{\text{spin}} \int d\Pi(q,l) \left(\includegraphics[align=c, scale=1]{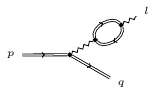}\right)\left(\includegraphics[align=c, scale=1]{compton/nonlinear_compton_lab.pdf}\right)^*.
\end{split}
\end{equation} 
Here, there are no identical particles in the intermediate state, as well as anti-particles put on a mass shell, and no fermion loop was cut. Hence, there are no additional factors in the RHS. Overall, the cutting equation for $\mT_{pp}^{(4),\mathrm{pol}}$ reads:
\begin{equation}
\label{eq:cutting_equation_two_loop_diagrams}
\begin{split}
2 \operatorname{Im} \left(\includegraphics[align=c, scale=1]{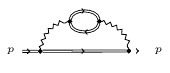}\right) = & \sum_{\text{spin}} \int d \Pi(q,g,f) \left(\left|\includegraphics[align=c, scale=1]{trident/trident_dir_lab.pdf}\right|^2 + \left|\includegraphics[align=c, scale=1]{trident/trident_exch_lab.pdf}\right|^2 \right) 
\\ & + 2 \operatorname{Re} \sum_{\text{spin}} \int d\Pi(q,l) \left(\includegraphics[align=c, scale=1]{radpol/radpol_lab.pdf}\right)\left(\includegraphics[align=c, scale=1]{compton/nonlinear_compton_lab.pdf}\right)^*.
\end{split}
\end{equation}

As mentioned in the previous section, for the cutting equation to make sense, we have to make sure that the LHS is gauge-invariant and diagrams on the RHS satisfy the Ward-Takahashi identity. Therefore, we can apply the cutting equation only to the sum of the remaining forward scattering diagrams. Cuts of the two diagrams, which put two photons on shell, together contribute to the double Compton scattering on the RHS:
\begin{equation}\label{eq:cut_2compton}
    \begin{split}
        \includegraphics[align=c, scale=1]{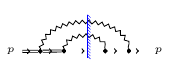} +& \includegraphics[align=c, scale=1]{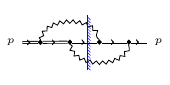} \\
        & = \sum_{\text{spin}} \int d\Pi(q,l,l') \left|\includegraphics[align=c, scale=1]{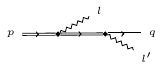} + \includegraphics[align=c, scale=1]{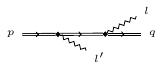}\right|^2.
    \end{split}
\end{equation}
One of the cuts of the third diagram contains the interference term for the trident process
\begin{equation}
    \begin{split}
        \includegraphics[align=c, scale=1]{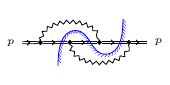} = -2 \sum_{\text{spin}} \int d\Pi(q,g,f) \left(\includegraphics[align=c, scale=1]{trident/trident_dir_lab.pdf}\right)\left(\includegraphics[align=c, scale=1]{trident/trident_exch_lab.pdf}\right)^*.
    \end{split}
\end{equation}
Here, no fermion loop is cut, and the $(-1)$ factor comes from the on-shell intermediate positron. The remaining cuts contain radiative corrections to the nonlinear Compton effect:
\begin{equation}
    \begin{split}
        \includegraphics[align=c, scale=1]{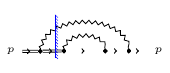} & + \includegraphics[align=c, scale=1]{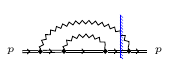} \\
        & = 2\operatorname{Re}\sum_{\text{spin}} \int d\Pi(q,l) \left(\includegraphics[align=c, scale=1]{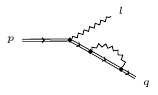}\right)\left(\includegraphics[align=c, scale=1]{compton/nonlinear_compton_lab.pdf}\right)^*,
    \end{split}
\end{equation}
\begin{equation}
    \begin{split}
        \includegraphics[align=c, scale=1]{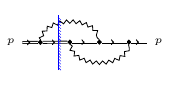} & + \includegraphics[align=c, scale=1]{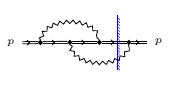} \\
        & = 2\operatorname{Re}\sum_{\text{spin}} \int d\Pi(q,l) \left(\includegraphics[align=c, scale=1]{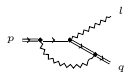}\right)\left(\includegraphics[align=c, scale=1]{compton/nonlinear_compton_lab.pdf}\right)^*.
    \end{split}
\end{equation}
Overall, the corresponding cutting equation reads:
\begin{equation}\label{eq:cutting_equation_two_loop_diagrams_1}
    \begin{split}
        &2 \operatorname{Im}\Bigg(\includegraphics[align=c, scale=1]{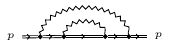} + \includegraphics[align=c, scale=1]{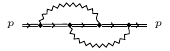}\Bigg) \\
        & \quad= \sum_{\text{spin}} \int d\Pi(q,l,l') \left|\includegraphics[align=c, scale=1]{2compton/2compton_dir_lab.pdf} + \includegraphics[align=c, scale=1]{2compton/2compton_exch_lab.pdf}\right|^2 \\
        &\quad\quad + 2\operatorname{Re}\sum_{\text{spin}} \int d\Pi(q,l) \left(\includegraphics[align=c, scale=1]{vertexpol/vertexpol_lab.pdf} + \includegraphics[align=c, scale=1]{scatterpol/scatterpol_lab.pdf} \right)\left(\includegraphics[align=c, scale=1]{compton/nonlinear_compton_lab.pdf}\right)^* \\
        &\quad\quad -2\sum_{\text{spin}} \int d\Pi(q,g,f) \left(\includegraphics[align=c, scale=1]{trident/trident_dir_lab.pdf}\right)\left(\includegraphics[align=c, scale=1]{trident/trident_exch_lab.pdf}\right)^*.
    \end{split}
\end{equation}

By combining eqs.~\eqref{eq:cutting_equation_two_loop_diagrams} and \eqref{eq:cutting_equation_two_loop_diagrams_1}, we arrive at:
\begin{equation}\label{eq:cutting_equation_two_loop_diagrams_2}
    \begin{split}
        &2\operatorname{Im}\mT_{pp}^{(4)} = \sum_{\text{spin}} \int d\Pi(q,l,l') \left|\includegraphics[align=c, scale=1]{2compton/2compton_dir_lab.pdf} + \includegraphics[align=c, scale=1]{2compton/2compton_exch_lab.pdf}\right|^2 \\
        & + \sum_{\text{spin}} \int d\Pi(q,g,f) \left|\includegraphics[align=c, scale=1]{trident/trident_dir_lab.pdf} - \includegraphics[align=c, scale=1]{trident/trident_exch_lab.pdf}\right|^2 \\
        & + 2\operatorname{Re}\sum_{\text{spin}} \int d\Pi(q,l) \left(\includegraphics[align=c, scale=1]{vertexpol/vertexpol_lab.pdf}\right)\left(\includegraphics[align=c, scale=1]{compton/nonlinear_compton_lab.pdf}\right)^* \\
        & + 2\operatorname{Re}\sum_{\text{spin}} \int d\Pi(q,l) \left(\includegraphics[align=c, scale=1]{scatterpol/scatterpol_lab.pdf} + \includegraphics[align=c, scale=1]{radpol/radpol_lab.pdf} \right)\left(\includegraphics[align=c, scale=1]{compton/nonlinear_compton_lab.pdf}\right)^*
    \end{split}
\end{equation}
According to eq.~\eqref{eq:s_matrix_sfqed}, the matrix element $\mT_{pp}^{(4)}$ is proportional to the large space-time volume $VT$. By expressing it as $\mT_{pp}^{(4)} = \mathcal{M}^{(4)}(p) VT$ and using the relation between the $\mT$-matrix and rates:
\begin{equation}\label{eq:optical_theorem_2_loop}
    W_X = \frac{1}{T}\frac{1}{N_X} \sum_{\text{spin}}\int \prod_{k=1}^{n_e(X)} \frac{d^3 \bm{p}_k V}{(2\pi)^3} \prod_{k=1}^{n_\gamma(X)} \frac{d^3 \bm{l}_k V}{(2\pi)^3} \frac{|\mT_{Xp}|^2}{\braket{X}\braket{p}} = \frac{1}{2p^0 VT} \sum_{\text{spin}} \int d\Pi_X |\mT_{Xp}|^2 ,
\end{equation}
the cutting equation \eqref{eq:cutting_equation_two_loop_diagrams_2} can be cast into the form:
\begin{equation}
    2\operatorname{Im}\mathcal{M}^{(4)}(p) = 2p^0 \left(W^{(4)}_{\gamma\gamma e^-} + W^{(4)}_{e^- e^+ e^-} + W^{(4)}_{\gamma e^-,\text{vert.}} + W^{(4)}_{\gamma e^-,\text{out}}\right),
\end{equation}
which resembles the optical theorem:
\begin{equation}
    2\operatorname{Im} \mathcal{M}(p) = 2p^0 \sum_{X} W_X.
\end{equation}
Here, $W^{(4)}_{\gamma\gamma e^-}$ and $W^{(4)}_{e^- e^+ e^-}$ are the rates for the double photon emission and the trident pair production at the lowest order, respectively. The remaining terms in the RHS, namely $W^{(4)}_{\gamma e^-,\text{vert.}}$ and $W^{(4)}_{\gamma e^-,\text{out}}$, contain radiative corrections to the photon emission process. The former corresponds to the vertex and the latter to the outgoing states' radiative corrections.

Let us emphasize that, while $W^{(4)}_{\gamma\gamma e^-}, W^{(4)}_{e^- e^+ e^-}$ and $W^{(4)}_{\gamma e^-,\text{vert.}}$ are expected in the RHS of the perturbative expansion of the optical theorem, the occurrence of  $W^{(4)}_{\gamma e^-,\text{out}}$ is non-trivial. It is common in background-free QFTs that loop corrections to in- and outgoing states are absorbed into real-valued renormalization constants, and only amputated diagrams are considered. However, in the presence of a background field, loop corrections gain field-induced complex-valued UV-finite parts (see, e.g. \cite{ritus1972radiative, mironov2022structure}) that cannot be absorbed into a renormalization constant without breaking the hermiticity of the Hamiltonian. Of course, this is only a brief look at the problem. A deeper understanding of the nature of these extra terms requires a more thorough study of the renormalization procedure and reduction formalism in a background field, which is out of the scope of this study. At higher orders in perturbation theory, more terms containing radiative corrections to the outgoing states would emerge in the RHS of the optical theorem, with the cutting equation providing a systematic way of identifying such terms.

Our next goal is to check eq.~\eqref{eq:cutting_equation_two_loop_diagrams} explicitly in a special case of constant crossed field and apply it to a trident process. To this end, in section~\ref{sec:v}, we consider explicitly and in greater detail the cutting of the two-loop polarization correction to the electron elastic scattering amplitude and in section~\ref{sec:vi}, we derive an expression for the direct contribution to the probability of the trident process directly from the Feynman rules.

\section{Cutting the two-loop polarization correction to the electron elastic scattering}
\label{sec:v}
In this section, we give a closer look at $\mT_{pp}^{(4),\mathrm{pol}}$, see the first diagram in eq.~\eqref{eq:T4_diagrams}. Equation~\eqref{eq:cutting_equation_two_loop_explicit_amplitude} illustrates how cutting equations apply to $\mT_{pp}^{(4),\mathrm{pol}}$ \textit{at the level of the matrix element definitions}. It is natural to ask whether the information about the trident process and the polarization correction to nonlinear Compton emission can be extracted from the \textit{integrated} two-loop scattering amplitude, particularly, its \textit{imaginary part}. Such amplitude was obtained by Ritus in ref.~\cite{ritus1972radiative} and later generalised in refs.~\cite{mironov2020resummation,mironov2022structure}. In ref.~\cite{ritus1972vacuum}, Ritus suggests ad hoc cutting rules to extract the mentioned rates. We reconsider this amplitude and apply the cutting rules formulated in previous sections, and explicitly show how the optical theorem applies in this example.

\subsection{Extracting the two-loop correction from the resummed bubble-chain amplitude}
\label{sec:v_1}
Let us consider elastic scattering of an electron in a CCF with initial momentum $p^\mu$, and characterised by the quantum dynamical parameter\footnote{In what follows, the $\chi$ parameter of secondary particles (real or virtual) is defined similarly by replacing $p$ in the defining eq.~\eqref{eq:chi_def} and in the subscript with the desired momentum (e.g., $\chi_q$ for a particle with momentum $q$).}
\begin{equation}
    \label{eq:chi_def}
    \chi_p=\frac{|e|}{m^3}\sqrt{-(\mF^{\mu\nu}p_\nu)^2}.
\end{equation}
where $\mF^{\mu\nu}$ is the  electromagnetic background field tensor. For brevity, for the incoming electron, we denote $\chi=\chi_p$ throughout.

The polarization contribution to the scattering matrix at two loops $\mT_{pp}^{(4),\mathrm{pol}}$, shown in the first diagram in eq.~\eqref{eq:T4_diagrams}, can be extracted from the resummed bubble-chain result $\mT_{pp}^{\text{b-c}}=\mM^{\text{b-c}}(\chi)VT$,\footnote{Note that the mass operator in a CCF is diagonal in $E_p$-representation \cite{ritus1972radiative}, namely, the total momentum is conserved, as it trivially follows from eq.~\eqref{eq:s_matrix_sfqed}.} where the amplitude
$\mM^{\text{b-c}}(\chi)=-\sum_{i=0}^{2} \mM_i(\chi)$ was derived in refs.~\cite{mironov2020resummation, mironov2022structure}:\footnote{The minus sign in the definition of $\mM_i$ is introduced to match the conventions of  refs.~\cite{mironov2020resummation,mironov2022structure}.} 
	\begin{eqnarray}
		\label{M0}
		\begin{aligned}
			\mM_0(\chi)=\frac{\alpha m^2}{(2\pi)^2} \int_{-\infty}^{\infty} & \frac{du}{(1+u)^2}\, \int_{-\infty}^{\infty} d l^2 \int_{-\infty}^{\infty} \frac{d\mu}{\mu+i0}\,D_0(l^2,\chi_l)  \\
			&\times \left[\left(2+\frac{l^2}{m^2}-\frac{\mu}{m^2}\right) \Ai_1(t) + 2\frac{u^2+2u+2}{1+u}\left(\frac{\chi}{u}\right)^{2/3}\Ai'(t) 
            \vphantom{\left(\frac{u}{\chi}\right)^{2/3}}\right. \\
            &\left. \quad\quad -\frac{4\gamma_s}{(1+u)} \left(\frac{u}{\chi}\right)^{2/3} \Ai(t)  \right],
		\end{aligned}
	\end{eqnarray}
	\begin{equation}
		\label{M12}
		\begin{split}
			\mM_{1,2}(\chi)=&-\frac{\alpha m^2}{(2\pi)^2}\int_{-\infty}^{\infty}\frac{du}{(1+u)^2}\,\int_{-\infty}^{\infty} dl^2 \int_{-\infty}^{\infty} \frac{d\mu}{\mu+i0}\,   D_{1,2}(l^2,\chi_l)\\
			&\times\left[ \vphantom{\left(\frac{u}{\chi}\right)^{2/3}} \left( 1+\frac{l^2}{m^2}\frac{u^2+2u+2}{2u^2}-\frac{\mu}{m^2}\frac{u(2+u)}{2u^2}\right){\rm Ai}_1(t) \right.\\
             & \left.+ \left(\frac{u^2+2u+2}{1+u}\pm 1\right)\left(\frac{\chi}{u}\right)^{2/3}{\rm Ai}'(t) -2\gamma_s\left(\frac{1}{1+u}\pm 1 \right) \left(\frac{u}{\chi}\right)^{2/3} {\rm Ai}(t)  \right].
		\end{split}
	\end{equation}
Here, $\alpha=e^2/4\pi$ is the fine structure constant,
\begin{eqnarray}
	\label{t}
	t=\left(\frac{u}{\chi}\right)^{2/3}\left(1+\frac{1+u}{u^2}\frac{l^2}{m^2} +\frac{1+u}{u}\frac{\mu}{m^2}\right),\\
	\label{chi_l}
	\chi_l=\frac{u\chi }{1+u},
\end{eqnarray}
and $\mu=q^2-m^2$. The spin terms depend on  $\gamma_s=|e| \mF^\star_{\mu\nu}p^\mu s^\nu /2m^3$, where $\mF^\star_{\mu\nu}=\frac{1}{2}\varepsilon_{\mu\nu\rho\sigma}\mathcal{F}^{\rho\sigma}$ is the dual electromagnetic tensor and $s^\nu=\bar{u}_{p,\lambda} \gamma^\nu\gamma^5 u_{p,\lambda}/2m$ is the electron spin 4-vector \cite{berestetskii1982quantum, mironov2020resummation}. The Airy functions are defined by their integral representation:
\begin{equation}
    \label{eq:airy_def}
	\Ai(t)=\frac{1}{2\pi}\int_{-\infty}^\infty d\sigma \, e^{-i\sigma^3/3-it \sigma},\quad \Ai_1(t)=\frac{-i}{2\pi}\int_{-\infty}^\infty \frac{d\sigma}{\sigma-i\varepsilon} \, e^{-i\sigma^3/3-it \sigma}.
\end{equation}
The all-order resummed bubble-chain photon propagator used in eqs.~\eqref{M0}, \eqref{M12} reads:\footnote{We omit the gauge-dependent term, as the polarization correction to the mass operator and the related amplitude are gauge-invariant.}
\begin{equation}
	\label{D_bubble_chain}
	\begin{split}
		D^c_{\mu\nu}(l)= & D_0(l^2,\chi_l) \left(g_{\mu\nu}-\frac{l_\mu l_\nu}{l^2}\right)+\sum\limits_{i=1}^2 D_{i}(l^2,\chi_l) \epsilon_\mu^{(i)}(l)\epsilon_\nu^{(i)}(l),
	\end{split}
\end{equation}
where vectors $\epsilon_\mu^{(1)}(l)=|e|\mF_{\mu\nu}l^\nu/(m^3\chi_l)$ and $\epsilon_\mu^{(2)}(l)=|e|\mF^\star_{\mu\nu}l^\nu/(m^3\chi_l)$ define the photon polarization axes. The scalar functions are expressed as
\begin{equation}
D_0(l^2,\chi_l)=\frac{-i}{l^2-\Pi_0},
\end{equation}
\begin{equation}
	\begin{split}
			D_{1,2}(l^2,\chi_l)&=\frac{i\Pi_{1,2}}{(l^2-\Pi_0)(l^2-\Pi_0-\Pi_{1,2})}=\frac{-i}{l^2-\Pi_0}-\frac{-i}{l^2-\Pi_0-\Pi_{1,2}}.
		\end{split}
\end{equation}
The shorthands $\Pi_0\equiv l^2\widehat{\Pi}$ and $\Pi_{1,2}$ denote the 1-particle irreducible photon polarization operator eigenfunctions in a CCF:
\begin{equation}\label{polop_struct}
	\Pi_{\mu\nu}(l)= \Pi_0(l^2,\chi_l) \left(g_{\mu\nu}-\frac{l_\mu l_\nu}{l^2}\right)+\sum\limits_{i=1}^2\Pi_i(l^2,\chi_l)\epsilon_\mu^{(i)}(l)\epsilon_\nu^{(i)}(l).
\end{equation}

In the current work, for the bubble chain, we consider the renormalised one-loop polarization insertions. The corresponding $\Pi_{\mu\nu}(l)$ eigenfunctions read \cite{ritus1972radiative}:
\begin{equation}\label{pi0}
		\Pi_0(l^2,\chi_l)=-l^2\frac{4\alpha}{\pi}  \int_4^\infty \frac{dv}{v^{5/2}\sqrt{v-4}} \left[f_1(\zeta)-\log\left(1-\frac{1}{v}\frac{l^2}{m^2}\right)\right],
\end{equation}
\begin{equation}\label{pi12}
	\Pi_{1,2}(l^2,\chi_l)=\frac{4\alpha\chi_l^{2/3} m^2}{3\pi}\int_4^\infty \frac{dv}{v^{13/6}} \,\frac{v+0.5\mp1.5}{\sqrt{v-4}}f'(\zeta).
\end{equation}
Here,
\begin{equation}\label{zeta}
	\zeta=\left(\frac{v}{\chi_l}\right)^{2/3}\left(1-\frac{l^2}{vm^2}\right)
\end{equation}
is the argument of the Ritus's functions
\begin{equation}\label{ritus_functions}
	\begin{aligned}
		f(\zeta)=i\int_0^\infty d\sigma\, e^{-i(\zeta\sigma+\sigma^3/3)},\\
		f_1(\zeta)=\int_{\zeta}^\infty dz \left[f(z)-\frac{1}{z}\right]=\int_0^\infty \frac{d\sigma}{\sigma} e^{-i\zeta\sigma} \left(e^{-i\sigma^3/3}-1\right).
	\end{aligned}
\end{equation}
In what follows, we use the property $\Im f(\zeta)=\pi\mathrm{\Ai}(\zeta)$ \cite{ritus1985quantum}. Note that $\Im f_1(\zeta)=\pi \Ai_1^{\rm (reg)}$, where
\begin{equation}
	\Ai^{\mathrm{(reg)}}_1(t)=\frac{-i}{2\pi}\int_{-\infty}^\infty \frac{d\sigma}{\sigma} \, e^{-it \sigma}\left(e^{-i\sigma^3/3}-1\right),
\end{equation}
differs from the $\Ai_1$ function entering eqs.~\eqref{M0}, \eqref{M12}. The regularised and nonregularised $\Ai_1$ functions differ by $\Ai_1(\zeta)- \Ai^{\mathrm{(reg)}}_1(\zeta)=\theta(-\zeta)$. 

To obtain the two-loop correction induced by a single bubble insertion $\mT_{pp}^{(4),\mathrm{pol}}$, we expand $D_i$ in $\alpha$:
\begin{equation}
	D_0(l^2,\chi_l)= \frac{-i}{l^2+i\varepsilon}+\frac{-i}{l^2+i\varepsilon}i\Pi_0\frac{-i}{l^2+i\varepsilon} +O(\alpha^2),
\end{equation}
\begin{equation}\label{photon_prop_trans}
	\begin{split}
		D_{1,2}(l^2,\chi_l)=-\frac{-i}{l^2+i\varepsilon}i\Pi_{1,2}\frac{-i}{l^2+i\varepsilon}+O(\alpha^2).
	\end{split}
\end{equation}
In order $O(\alpha)$, the polarization eigenvalues are multiplied by $((-i)/(l^2+i\varepsilon))^2$, which corresponds to two photon propagators that carry the incoming and outgoing momentum $l$.
By substituting these terms into eqs.~\eqref{M0} and \eqref{M12}, we arrive at
	\begin{equation}
		\label{M_two_loop}
		\begin{split}
			\mM_{i}^{(4)}(\chi)=&-\frac{\alpha m^2}{(2\pi)^2}\int_{-\infty}^{\infty}\frac{du}{(1+u)^2}\,\int_{-\infty}^{\infty} dl^2 \int_{-\infty}^{\infty} d\mu \,  \frac{i\Pi_{i}(l^2,\chi_l)}{(\mu+i\varepsilon)(l^2+i\varepsilon)^2} \mathcal{R}_{i}(\mu,l^2,u),
		\end{split}
	\end{equation}
where $i=0,1,2$. By real functions $\mathcal{R}_i(\mu,\lambda,u)$ we denote the expressions contained in the square brackets in eqs.~\eqref{M0} and \eqref{M12}. Hence, the corresponding matrix element reads:
\begin{equation}
    \mT_{pp}^{(4),\mathrm{pol}}(\chi)=-VT\sum_i \mM_{i}^{(4)}(\chi).
\end{equation}

The $\mM_0$ component in eq.~\eqref{M0} (as well as $\mM_0^{(4)}$) contains a divergent vacuum contribution. As our goal is to identify the trident process rate contribution to $\Im \mT_{pp}^{(4),\mathrm{pol}}$, which vanishes at $\mF=0$, we will focus on the finite field-dependent contribution. It is handy to regularise amplitude $\mM$ by subtracting its value at vanishing field, $\mM^{\rm (reg)}=\mM-\mM |_{\mF=0}$. Thus, from now on, we will consider the regularised version of eq.~\eqref{M_two_loop}.

To identify the imaginary part of $\mM^{(4)}$, let us consider the following structure under the integral in eq.~\eqref{M_two_loop}:
\begin{equation}
    \label{combination}
	\frac{i\Pi_i(\lambda,\chi_l)}{(\mu+i\varepsilon)(\lambda+i\varepsilon)^2}.
\end{equation}
Here, for brevity, we introduced $\lambda=l^2$. It is straightforward to split eq.~\eqref{combination} into the real and imaginary parts. The denominator regularized with the $\varepsilon$-prescription is interpreted as a distribution (in the sense of generalized functions). On the other hand, we can view eq.~\eqref{combination} as a combination of multiple propagators in the momentum space, and apply the cutting rules, as formulated in section~\ref{sec:iii}. Let us discuss both approaches and compare them.

\subsection{Cutting the two-loop amplitude $\mT_{pp}^{(4),\mathrm{pol}}$}\label{sec:v_2}
The result for $\mT_{pp}^{(4),\mathrm{pol}}$, expressed via eq.~\eqref{photon_prop_trans}, is formulated in the $E_p$-representation \cite{mironov2020resummation}. Therefore, to perform a unitarity cut, we can apply the rules in the $p$-space as formulated in eqs.~\eqref{eq:cutting_rules_momentum_1}, \eqref{eq:cutting_rules_momentum_2}. This will allow us to extract the trident pair production rate and the polarization correction to the nonlinear Compton emission rate [the first and second terms in the RHS of eq.~\eqref{eq:cutting_equation_two_loop_diagrams},  respectively]. Ritus suggests a similar set of rules in ref.~\cite{ritus1972vacuum}. Therefore, let us review these rules and compare them to our formulation.

First, consider the central cut of the diagram, as shown in eq.~\eqref{eq:cut_trident}, which is associated with the direct part of the trident process. According to ref.~\cite{ritus1972vacuum}, the following cutting rules apply: 
\begin{enumerate}
	\item Cut the electron propagator carrying the momentum $q$ by the substitution: \\$\dfrac{1}{\mu+i\varepsilon}\mapsto -2i \pi \delta(\mu)\theta(q_-)$;
	\item Cut the photon polarization operator by $\Pi_i\mapsto -2i\, \theta(l_-)\,\Im\Pi_i$;
	\item Replace the photon propagator that is on the RHS of the cut by its complex conjugate: $(-i)/(\lambda+i\varepsilon)\mapsto i/(\lambda-i\varepsilon)$.
\end{enumerate}
The first and third rules match our eqs.~\eqref{eq:cutting_rules_momentum_1}, \eqref{eq:cutting_rules_momentum_2}, whereas the second rule follows from the cutting equation applied to the one-loop polarization operator itself.  Let us show the latter explicitly. In the coordinate space, the polarization operator is given by
\begin{equation}
    i\Pi_{\mu\nu}(x,y) = (-1)\operatorname{tr}\left[(-ie\gamma^\mu)S(x,y)(-ie\gamma^\nu)S(y,x)\right] \equiv F_{\mu\nu}(x,y) .
\end{equation}
The cutting equation \eqref{eq:cutting_equation} reads
\begin{equation}
    F_{\mu\nu}(x,y) + F_{\mu\nu}^*(x,y) = -F_{\mu\nu}(y|x) -F_{\mu\nu}(x|y),
\end{equation}
where
\begin{align}
    & F_{\mu\nu}(x,y) + F_{\mu\nu}^*(x,y) = i \Pi_{\mu\nu}(x,y) - i \Pi^*_{\mu\nu}(x,y), \\
    & F_{\mu\nu}(y|x) =  (-1)\operatorname{tr}\left[(+ie\gamma^\mu)S^{(+)}(x,y)(-ie\gamma^\nu)S^{(-)}(y,x)\right], \\
    & F_{\mu\nu}(x|y) = (-1)\operatorname{tr}\left[(-ie\gamma^\mu)S^{(-)}(x,y)(+ie\gamma^\nu)S^{(+)}(y,x)\right].
\end{align}
In the momentum space, the polarization operator is diagonal:
\begin{equation}
    \Pi_{\mu\nu}(l) (2 \pi^4) \delta(l-l') = \int d^4x \, d^4y \, \Pi_{\mu\nu}(x,y) e^{ixl-iyl'}
\end{equation}
and is an even function of the momentum: $\Pi_{\mu\nu}(l) = \Pi_{\mu\nu}(-l)$ [see eq.~\eqref{polop_struct}]. It follows that
\begin{align}\label{eq:im_polop_to_cuts}
    2 \operatorname{Im} \big(\Pi_{\mu\nu}(l)\big) \,(2\pi)^4 \delta(l-l') = \int d^4x \, d^4y \,  \left[F_{\mu\nu}(y|x) + F_{\mu\nu}(x|y)\right] e^{ixl-iyl'} .
\end{align}
The positive-energy Wightman function $S^{(+)}(x,y)$ demands positive light-front momentum flow from the vertex $y$ to the vertex $x$. Conversely, the negative-energy Wightman function $S^{(-)}(x,y)$ demands positive light-front momentum flow from the vertex $x$ to the vertex $y$. Therefore
\begin{align}
    & \int d^4 x \, d^4 y \, F_{\mu\nu}(y|x) e^{ixl-iyl'} \propto \theta(l_-) \,(2\pi)^4 \delta(l-l'), \\
    & \int d^4 x \, d^4 y \, F_{\mu\nu}(x|y) e^{ixl-iyl'} \propto \theta(-l_-) \,(2\pi)^4 \delta(l-l').
\end{align}
By multiplying eq.~\eqref{eq:im_polop_to_cuts} by $\theta(l_-)$ we arrive at
\begin{equation}
    \int d^4 x \, d^4 y \, F_{\mu\nu}(y|x) e^{ixl-iyl'} =  2 \operatorname{Im} \big(\Pi_{\mu\nu}(l)\big) \, \theta(l_-) \,(2\pi)^4 \delta(l-l').
\end{equation}
A cut shown in eq.~\eqref{eq:cut_trident} corresponds to the substitution 
$$i\Pi_{\mu\nu}(x,y) \mapsto F_{\mu\nu}(y|x).$$ Therefore, in the momentum space, we indeed arrive at the substitution
\begin{equation}
    \Pi_{\mu\nu} \mapsto -2i \operatorname{Im} \Pi_{\mu\nu} \, \theta(l_-).
\end{equation}

By applying the listed rules to eq.~\eqref{combination}, we obtain:
\begin{equation}
    \label{substitution_trident}
    \left[\frac{i\Pi_i(\lambda,\chi_l)}{(\mu+i\varepsilon)(\lambda+i\varepsilon)^2}\right]^{\rm trident} \mapsto i 4\pi\theta(q_-)\theta(l_-)\delta(\mu) \Im\Pi_i\times \frac{1}{\lambda^2+\varepsilon^2}.
\end{equation}
The rightmost fractional factor can be rewritten as
\[
\frac{1}{\lambda^2+\varepsilon^2}=\frac{\lambda^2-\varepsilon^2}{[\lambda^2+\varepsilon^2]^2}+2\frac{\varepsilon^2}{[\lambda^2+\varepsilon^2]^2}.
\]
In the limit $\varepsilon\rightarrow 0$, the first term turns into the distribution $-\left(\frac{\mP}{\lambda}\right)'$, where $\mP$ denotes the Cauchy principal value.\footnote{This is straightforward to show by considering a derivative of the distribution $\mP/\lambda=\lim_{\varepsilon\rightarrow0} \lambda/(\lambda^2+\varepsilon^2)$.} Up to a numerical factor, the second term is shaped as the squared delta-function  $\delta(\lambda)=\lim_{\varepsilon\rightarrow0} \varepsilon/\pi (\lambda^2+\varepsilon^2)$, therefore we denote it by $2\pi^2\delta^2(\lambda)$. While this term is not a well-defined distribution, we will discuss its physical interpretation in what follows. Hence, the overall resulting substitution reads:
\begin{equation}
    \label{substitution_trident2}
    \left[\frac{i\Pi_i(\lambda,\chi_l)}{(\mu+i\varepsilon)(\lambda+i\varepsilon)^2}\right]^{\rm trident} \mapsto i 4\pi\theta(q_-)\theta(l_-)\delta(\mu) \Im\Pi_i\left[-\left(\frac{\mP}{\lambda}\right)'+2\pi^2\delta^2(\lambda)\right].
\end{equation}

The contributions from the two other possible cuts, shown in eq.~\eqref{eq:cut_rad_corr}, should be summed and considered together. They are assigned to a correction to photon emission. For the cut going through the rightmost photon propagator, the rules are the following:
\begin{enumerate}
    \item Cut the electron propagator: $\dfrac{1}{\mu+i\varepsilon}\mapsto -2i \pi \delta(\mu)\theta(q_-)$;
    \item Cut the rightmost photon propagator in the outer loop: $\dfrac{1}{\lambda+i\varepsilon}\mapsto -2i \pi \delta(\lambda)\theta(l_-)$.
\end{enumerate}
For the second term in the LHS of eq.~\eqref{eq:cut_rad_corr}, the first rule remains the same, however, the others read:
\begin{enumerate}
    \item[2.*] Cut the leftmost photon propagator in the outer loop: $\dfrac{1}{\lambda+i\varepsilon}\mapsto -2i \pi \delta(\lambda)\theta(l_-)$;
	\item[3.*] Replace the photon propagator to the right of the cut by its complex conjugate: $(-i)/(\lambda+i\varepsilon)\mapsto i/(\lambda-i\varepsilon)$;
    \item[4.*] Replace the polarization operator by its complex conjugate: $i\Pi_i\mapsto (-i)\Pi_i^*$.
\end{enumerate}
As before, these rules follow directly from eqs.~\eqref{eq:cutting_rules_momentum_1}, \eqref{eq:cutting_rules_momentum_2} and coincide with suggested in ref.~\cite{ritus1972vacuum}. After applying the rules and summing the two contributions to eq.~\eqref{combination} for the two cuts, we obtain:
\begin{equation}
\begin{split}\label{substitution_rad_pol_initial}
    \left[\frac{i\Pi_i(\lambda,\chi_l)}{(\mu+i\varepsilon)(\lambda+i\varepsilon)^2}\right]_{\rm left}^{\rm rad\,pol}+&\left[\frac{i\Pi_i(\lambda,\chi_l)}{(\mu+i\varepsilon)(\lambda+i\varepsilon)^2}\right]_{\rm right}^{\rm rad\,pol}\\
    &\mapsto -i8\pi^2 \theta(q_-)\theta(l_-)\delta(\mu) \Re\left[ \frac{\delta(\lambda) \Pi_i}{\lambda+i\varepsilon} \right].
\end{split}
\end{equation}
Here, the arising distribution in $\lambda$ can be rewritten as follows \cite{ritus1972vacuum}:
\[
\frac{\pi\delta(\lambda)}{\lambda+i\varepsilon}=\frac{\lambda\varepsilon}{(\lambda^2+\varepsilon^2)^2}+i\frac{\varepsilon^2}{(\lambda^2+\varepsilon^2)^2}\overset{\varepsilon\rightarrow 0}{\rightarrow} -\frac{1}{2}\delta'(\lambda)+i \pi^2\delta^2(\lambda),
\]
where we substituted the regularised delta function $\delta(\lambda)\sim\varepsilon/\pi (\lambda^2+\varepsilon^2)$. The $\delta'(\lambda)$ term can be derived by applying the distributional derivative to  $\delta(\lambda)$, while the second term is interpreted as before in the discussion of eq.~\eqref{substitution_trident}. Overall, the polarization correction to photon emission is obtained by substituting
\begin{equation}
    \label{substitution_rad_pol}
    \left[\frac{i\Pi_i(\lambda,\chi_l)}{(\mu+i\varepsilon)(\lambda+i\varepsilon)^2}\right]^{\rm rad\,pol}
    \mapsto i4\pi^2 \theta(q_-)\theta(l_-)\delta(\mu) 
    \left[\pi\delta'(\lambda)\Re\Pi_i-2\pi^2\delta^2(\lambda)\Im\Pi_i\right].
\end{equation}

Let us note that the $2\pi^2\delta^2(\lambda)$ terms enter eqs.~\eqref{substitution_trident2} and \eqref{substitution_rad_pol} with opposite signs. Therefore, these terms cancel exactly when the full cutting equation \eqref{eq:cutting_equation_two_loop_diagrams} is considered, meaning that the whole equation is regular and well-defined.

In his original work \cite{ritus1972vacuum}, after introducing the cutting rules and applying the substitutions \eqref{substitution_trident2} and \eqref{substitution_rad_pol} to the analogue of our expression \eqref{M_two_loop}, Ritus proceeded to integrate the resulting expression over $\mu$ and $\lambda$. The latter requires a careful treatment of the Cauchy principal value terms and the distributional derivative (namely, integrating the relevant terms by parts in $\lambda$). The resulting expression, e.g. for the trident process rate [see eq. (19) in ref.~\cite{ritus1972vacuum}] takes a unique form, that differs from the expressions used in the modern literature \cite{king2013trident,king2018effect}. Below, we show a different way of treating the integrals, but let us first discuss the explicit evaluation of the imaginary part of eq.~\eqref{combination}.

\subsection{Imaginary part of $\mT_{pp}^{(4),\mathrm{pol}}$}
\label{sec:v_3}
Let us consider  the particular cutting equation~\eqref{eq:cutting_equation_two_loop_diagrams} in further detail. As prescribed, we can calculate the imaginary part of $\mT_{pp}^{(4),\mathrm{pol}}$ in the LHS directly from $\mM_i^{(4)}$ as given in eq.~\eqref{M_two_loop} and compare the result against the application of the cutting rules.

Let us evaluate the imaginary parts of  eq.~\eqref{combination} in the limit $\varepsilon\rightarrow0$. We use the Sokhotski-Plemelj formula  $\lim_{\varepsilon\rightarrow 0}1/(\mu+i\varepsilon)=\mP/\mu-i\pi\delta(\mu)$. The limit of $1/(\lambda+i\varepsilon)^2$ is obtained by taking the distributional derivative of $1/(\lambda+i\varepsilon)$, which gives 
\[
\lim\limits_{\varepsilon\rightarrow 0}\frac{1}{(\lambda+i\varepsilon)^2}=-\left(\frac{\mP}{\lambda}\right)'+i\pi\delta'(\lambda).
\]
Hence, we arrive at
\begin{equation}
    \label{combination_im}
    \begin{split}
    &\Im\frac{i\Pi_i(\lambda,\chi_l)}{(\mu+i\varepsilon)(\lambda+i\varepsilon)^2}\\
    &\quad=\pi\delta(\mu)\left[-\left(\frac{\mP}{\lambda}\right)'\Im\Pi_i+\pi\delta'(\lambda)\Re\Pi_i\right] -\pi \frac{\mP}{\mu}\delta'(\lambda)\Im\Pi_i - \frac{\mP}{\mu}\left(\frac{\mP}{\lambda}\right)'\Re\Pi_i.
    \end{split}
\end{equation}
In anticipation that this expression should coincide with eqs.~\eqref{substitution_trident2} and \eqref{substitution_rad_pol} combined, we note two key differences. First, the last two terms from eq.~\eqref{combination_im} do not appear in eqs.~\eqref{substitution_trident2} and \eqref{substitution_rad_pol}. Second, $\theta$-functions $\theta(q_-)\theta(l_-)$ are not formed in the prefactor. It turns out that both issues are resolved by integrating eq.~\eqref{M_two_loop} by parts over $\mu$ and $\lambda$. 

To this end, consider the following auxiliary expression:
\[
J=\int_{-\infty}^\infty d\lambda \int_{-\infty}^\infty d\mu \frac{1}{[\lambda\pm i\sigma(l_-)\varepsilon]^2} \frac{1}{\mu\mp i\sigma(q_-)\varepsilon}\left[\Re\Pi_i(\lambda)+i\sigma(l_-)\Im\Pi_i(\lambda)\right]\Ai(z_1+a\mu+b\lambda),
\]
where $\sigma(l_-)$ and $\sigma(q_-)$ denote the sign functions, and the argument of the Airy function abbreviates eq.~\eqref{t}. The factor $1/[\mu\mp i\sigma(q_-)\varepsilon]$ can be viewed as the advanced/retarded propagator $D_{\rm adv,ret}(\mu)$, whilst $1/[\lambda\pm i\sigma(l_-)\varepsilon]$ as $D_{\rm ret,adv}(\lambda)$. Therefore, $J=0$ as a contraction of the advanced and retarded propagators. This can be shown by an explicit calculation using the integral representation of the Airy function and the fact that $\Pi_i(\lambda)$ is an entire function \cite{ritus1972radiative}. Indeed, by passing from $\mu$ and $\lambda$ to the light-front variables $q_+$ and $l_+$, respectively,\footnote{Recall that $\lambda=l^2=2l_+l_--l_\perp^2$, $\mu=q^2-m^2=2q_+q_--q_\perp^2-m^2$.} it is straightforward to show that 
\[
\int_{-\infty}^\infty d\mu \frac{e^{-ia\mu \sigma}}{\mu\mp i\sigma(q_-)\varepsilon}\propto \theta(\mp u\sigma),\quad \int_{-\infty}^\infty d\lambda\, \Pi_i(\lambda)\frac{e^{-ib\lambda \sigma}}{(\lambda\pm i\sigma(q_-)\varepsilon)^2}\propto \theta(\pm u\sigma),
\]
where $\sigma$ is the inner integration variable as defined in eq.~\eqref{eq:airy_def}. Hence, the product vanishes, as well as $J$. 

As a corollary, the replacement of the term given in eq.~\eqref{combination} with 
\begin{equation}
    \label{combination_with_sigma}
   \frac{\Re\Pi_i(\lambda)+i\sigma(l_-)\Im\Pi_i(\lambda)}{(\mu\mp i\sigma(q_-)\varepsilon)(\lambda\pm i\sigma(l_-)\varepsilon)^2}
\end{equation}
in the integral \eqref{M_two_loop} turns the latter to zero. Therefore, we can subtract the resulting expression from the initial eq.~\eqref{M_two_loop} without changing the total result. This is actually equivalent to properly taking the initial integral by parts. Let us use this handy property. Consider the imaginary part of eq.~\eqref{combination_with_sigma}:
\begin{equation}
    \label{combination_with_sigma_im}
    \begin{split}
    &\Im \frac{\Re\Pi_i(\lambda)+i\sigma(l_-)\Im\Pi_i(\lambda)}{(\mu\mp i\sigma(q_-)\varepsilon)(\lambda\pm i\sigma(l_-)\varepsilon)^2}\\
    &\quad =-\sigma(q_-)\sigma(l_-)\pi\delta(\mu)\left[-\left(\frac{\mP}{\lambda}\right)'\Im\Pi_i+\pi\delta'(\lambda)\Re\Pi_i \right]-\pi \frac{\mP}{\mu}\delta'(\lambda)\Im\Pi_i - \frac{\mP}{\mu}\left(\frac{\mP}{\lambda}\right)'\Re\Pi_i.
    \end{split}
\end{equation}
Note that upon subtraction of this expression from  eq.~\eqref{combination_im}, the  last two terms cancel:
\begin{equation}
\begin{split}
\Im \left[\frac{i\Pi_i(\lambda,\chi_l)}{(\mu+i\varepsilon)(\lambda+i\varepsilon)^2}\right. -&\left.\frac{\Re\Pi_i(\lambda)+i\sigma(l_-)\Im\Pi_i(\lambda)}{(\mu\mp i\sigma(q_-)\varepsilon)(\lambda\pm i\sigma(l_-)\varepsilon)^2}\right]\\
&=2\pi\theta(q_-)\theta(l_-)\delta(\mu)\left[-\left(\frac{\mP}{\lambda}\right)'\Im\Pi_i+\pi\delta'(\lambda)\Re\Pi_i \right].
\end{split}
\end{equation}
Here, we recover the $\theta$-functions as $[1+\sigma(q_-)\sigma(l_-)]=2\theta(q_-)\theta(l_-)$ (given that for the initial electron $p_->0$). Finally, the LHS of the cutting equation \eqref{eq:cutting_equation_two_loop_diagrams} can be calculated by evaluating
	\begin{equation}
		\label{twoImM_two_loop}
		\begin{split}
			2\Im\mM_{i}^{(4)}(\chi)=-\frac{\alpha m^2}{(2\pi)^2}&\int_{-\infty}^{\infty}\frac{du}{(1+u)^2} \,\int_{-\infty}^{\infty} d\lambda \int_{-\infty}^{\infty} d\mu  \\ &\times4\pi\theta(q_-)\theta(l_-)\delta(\mu)\left[-\left(\frac{\mP}{\lambda}\right)'\Im\Pi_i+\pi\delta'(\lambda)\Re\Pi_i \right] \mathcal{R}_{i}(\mu,l^2,u).
		\end{split}
	\end{equation}
We can see that the RHS of eq.~ \eqref{twoImM_two_loop} indeed coincides with the sum of the matrix elements obtained via the cutting prescriptions given in eqs.~\eqref{substitution_trident2} and \eqref{substitution_rad_pol}. Let us emphasise that the extra terms containing $\pm2\pi^2\delta^2(\lambda)$, which enter eqs.~\eqref{substitution_trident2} and \eqref{substitution_rad_pol}, do not show up in eq.~\eqref{twoImM_two_loop}.

\section{Direct contribution to trident process}\label{sec:vi}
In this section, we evaluate the direct contribution to the rate of the trident pair production. The corresponding diagram is shown in the RHS of eq.~\eqref{eq:cut_trident}. We perform this computation using two approaches and compare the results. First, we perform a direct tree-level calculation using momentum-space Feynman rules introduced in section~\ref{sec:iii_1}. We discuss the key intermediate steps and provide the full derivation in our computer-algebraic script available online \cite{github}, developed with the aid of \texttt{FeynCalc} package \cite{mertig1991feyn, shtabovenko2020feyncalc,shtabovenko2025feyncalc}. Second, we extract the rate from the two-loop polarization correction to the electron forward scattering using the cutting rules, as discussed in section~\ref{sec:v_2}.

\subsection{Tree-level calculation}
\label{sec:vi_1}
\subsubsection{Amplitude and invariant integration variables}
The partial amplitude for the direct diagram reads
\begin{equation}\label{eq:trident_amplitude}
    \mathcal{M}(s_ 1, s_ 2) = \left.\frac{e^2}{l^2+i0}\right|_{l=p-q+s_1 k} \, \overline{u}(\bm{g}) \, \Gamma ^{\mu} (s_ 2; g, -f) \, v(\bm{f}) \times \overline{u}(\bm{q}) \, \Gamma_{\mu}(s_ 1 ; q, p) \, u(\bm{p}),
\end{equation}
where the dressed vertex $\Gamma^\mu$ is given by eq.~\eqref{eq:Gamma_cn}.
The $\mT$-matrix element is obtained by integrating the partial amplitude multiplied by the momentum-conserving $\delta$-function over the momentum absorption numbers $s_1$ and $s_2$  [see eq.~\eqref{eq:s_matrix_sfqed}]:
\begin{equation}\label{eq:trident_s_matrix}
    \mT_{fi} = \int ds_1 ds_2 \, \mathcal{M}(s_1, s_2) \, (2\pi)^4 \delta\big(g+f+q-p-(s_1+s_2)k\big).
\end{equation}
For the sake of simplicity, we perform the calculation assuming the initial electron is unpolarized. Mod-squared $\mT$-matrix element, summed over the final and averaged over the initial particle spin states, is then given by\footnote{Note that, in contrast to previous occurrences, $\sum_{\text{spin}}$ in eq.~\eqref{eq:trident_s_matrix_squared} now also involves sum over the initial particle spin states.} 
\begin{align} \label{eq:trident_s_matrix_squared}
    \frac{1}{2}\sum_{\text{spin}}|\mT_{fi}|^2 =& \frac{VT}{2L_{\varphi}} \sum_{\text{spin}} \int ds_1 ds_2 dt_1 dt_2 \, \mathcal{M}(s_1, s_2) \mathcal{M}^*(t_1, t_2) \, (2\pi)^5 \delta(s_1+s_2-t_1-t_2) \nonumber \\ &\times \delta\big(g+f+q-p-(s_1+s_2)k\big),
\end{align}
where we have used eq.~\eqref{eq:delta_funcs_product} for the product of the delta-functions. The probability is expressed through the $\mT$-matrix by
\begin{equation}\label{eq:trident_prob}
    dP_{\text{dir}} = \frac{ \frac{1}{2}\sum_{\text{spin}}|\mT_{fi}|^2}{\langle i|i \rangle \langle f|f \rangle} \frac{V d^3\bm{q}}{(2\pi)^3} \frac{V d^3\bm{g}}{(2\pi)^3} \frac{V d^3\bm{f}}{(2\pi)^3} = \frac{\frac{1}{2}\sum_{\text{spin}}|\mT_{fi}|^2}{2p^0 V } \frac{d^3 \bm{q}}{2q^0(2\pi)^3} \frac{d^3 \bm{g}}{2g^0(2\pi)^3} \frac{d^3 \bm{f}}{2f^0(2\pi)^3}.
\end{equation}

Two electrons in the final state are indistinguishable; thus, the total probability should be divided by two after integrating over the whole phase volume. By combining the probability density \eqref{eq:trident_prob} with eq. \eqref{eq:trident_s_matrix_squared}, for the rate, we obtain:
\begin{align}\label{eq:trident_prob_rate0}
    W_{\text{dir}} = \frac{P_{\text{dir}}}{T} =& \sum_{\text{spin}} \int \frac{1}{64 p^0 q^0 g^0 f^0 (2\pi)^4 L_\varphi} \, \mathcal{M}(s_1, s_2) \mathcal{M}^*(t_1, t_2) \,\delta(s_1+s_2-t_1-t_2) \nonumber \\ &\times \delta\big(g+f+q-p-(s_1+s_2)k\big)\,ds_1 \,ds_2\, dt_1 \,dt_2\, d^3\bm{q} \, d^3\bm{g} \, d^3\bm{f},
\end{align}
where the spin sum is given by
\begin{align}\label{eq:spin_sum_def}
    &\sum_{\text{spin}}\mathcal{M}(s_1, s_2) \mathcal{M}^*(t_1, t_2) = \frac{e^4}{\big((p-q+s_1 k)^2+i0\big)\big((p-q+t_1 k)^2-i0\big)} \nonumber \\
    & \times \operatorname{tr}\big[ (\slashed{g}+m)\Gamma^\mu(s_2;g,-f)(\slashed{f}-m)\Gamma^{\nu*}(t_2;g,-f) \big] \operatorname{tr}\big[ (\slashed{q}+m)\Gamma_\mu(s_1;q,p)(\slashed{p}+m)\Gamma^{*}_\nu(t_1;q,p) \big].
\end{align}
The integral over the positron 3-momentum $\bm{f}$ is removed by the $\delta$-function as follows 
\begin{equation}
    \int d^3\bm{f} \, \left.\delta\left(g+f+q-p-(s_1+s_2)k\right)\right|_{f^0 = \sqrt{\bm{f}^2+m^2}} = \frac{f^0}{kf}\delta(s_1+s_2-\tilde{s}) = \frac{\xi f^0}{m^2 \chi_f} \delta(s_1+s_2-\tilde{s}),
\end{equation}
where $\tilde{s}$ is given by
\begin{equation}
    \tilde{s} = \frac{pq+pg-qg-m^2}{kf}.
\end{equation}
In the last step, we introduced dimensionless field strength $\xi=\sqrt{-e^2a^2}/m$ and used that in a CCF $\chi_f=\xi (kf)/m^2$. Integrals over $s_2$ and $t_2$ are removed by the two remaining $\delta$-functions, so that eq.~\eqref{eq:trident_prob_rate0} turns to 
\begin{align}\label{eq:trident_prob_rate1}
    W_{\text{dir}} = \sum_{\text{spin}} \int  \frac{\xi}{64 m^2 p^0 q^0 g^0 \chi_f (2\pi)^4 L_\varphi} \, \mathcal{M}(s, \tilde{s}-s) \mathcal{M}^*(t, \tilde{s}-t) \, ds \, dt \, d^3\bm{q} \, d^3\bm{g},
\end{align}
where we abbreviated $s =s_1$ and $t=t_1$. 
To evaluate the remaining integrals it is especially convenient to pass to the new integration  variables  suggested in ref.~\cite{ritus1985quantum}: 
\begin{equation}\label{eq:new_vars}
\begin{aligned}
    &u = \frac{\chi_l}{\chi_q},  &&\rho_1 = \frac{\alpha_{q,p}}{8 \beta_{q,p}} = \frac{e \mathcal{F}_{\mu\nu}q^\mu p^\nu}{\xi m^4 \chi_l},  &&\tau_1 = \frac{e \mathcal{F}^\star_{\mu\nu}q^\mu p^\nu}{ m^4 \chi_l}, \\
    &v = \frac{\chi_l^2}{\chi_g \chi_f},  &&\rho_2 = \frac{\alpha_{g,-f}}{8\beta_{g,-f}} = \frac{e \mathcal{F}_{\mu\nu}g^\mu f^\nu}{\xi m^4 \chi_l},  &&\tau_2 = \frac{e \mathcal{F}^\star_{\mu\nu}g^\mu f^\nu}{m^4 \chi_l} 
\end{aligned}
\end{equation}
As discussed in ref.~\cite{ritus1985quantum}, variables $\rho_1$ and $\rho_2$ represent the formation phase instants of the events of photon emission and pair photoproduction, respectively. The $\alpha$ and $\beta$  coefficients defined in appendix~\ref{sec:app_c} are expressed as
\begin{equation}
    \alpha_{q,p} = 8 \rho_1 \beta_{q,p}, \quad \alpha_{g,-f} = 8\rho_2 \beta_{g,-f},\quad
    \beta_{q,p} = \frac{\xi^3 u}{8 \chi}, \quad \beta_{g,-f} = \frac{\xi^3 (1+u)v}{8u\chi}.
\end{equation}
Variable $u$ defines quantum parameters of the initial electron and virtual photon,
\begin{equation}\label{eq:chi_ql}
     \chi_q = \frac{\chi}{1+u}, \quad\chi_l=\frac{u\chi}{1+u}.
\end{equation}
However, quantum parameters $\chi_g$ and $\chi_f$ of the created electron and positron are not uniquely fixed by variables $u$ and $v$. Rather, the solution has two branches
\begin{equation}
    \chi_g = \frac{\chi_l}{2}\left(1-\sigma\sqrt{1-\frac{4}{v}}\right), \quad \chi_f = \frac{\chi_l}{2}\left(1+\sigma\sqrt{1-\frac{4}{v}}\right)
\end{equation}
with $\sigma = +1$ and $\sigma = -1$. Therefore, integration over the phase volume of the outgoing electrons includes summation over these branches
\begin{equation}
    \int d^3 \bm{q} \, d^3 \bm{g} \longrightarrow \sum_{\sigma = \pm1} \int_0^\infty du \int_4^\infty dv \int_{-\infty}^{+\infty} d\tau_1\,d\tau_2\,d\rho_1\,d\rho_2 \, |J|,
\end{equation}
with Jacobian
\begin{equation}
    |J| = \frac{m^4 q^0 g^0 \xi^2  u^2 \chi_l}{(1+u)^3v\sqrt{v(v-4)} \, \chi_g}.
\end{equation}
Traces in eq.\eqref{eq:spin_sum_def} are expressed in terms of contractions of the encountered 4-vectors. In terms of new variables, all such
contractions can be rewritten as follows \cite{github}
\begin{eqnarray}
    aq = \frac{ap - e^{-1}m^2\xi^2\rho_1u}{1+u}, \\
    ag = \frac{u\chi_g}{(1+u)\chi_l}(ap+e^{-1}m^2\xi^2\rho_1)-e^{-1}m^2\xi^2\rho_2, \\
    af = \frac{u\chi_f}{(1+u)\chi_l}(ap+e^{-1}m^2\xi^2\rho_1)+e^{-1}m^2\xi^2\rho_2, \\
    pq = \frac{m^2 u^2 \left(\xi ^2 \rho _1^2+\tau _1^2+1\right)}{2 \left(1+u\right)}+m^2, \\
    fg = \frac{1}{2} m^2 \left(\frac{\chi _f}{\chi _g}+\frac{\chi _g}{\chi _f}\right) \left(\xi ^2 \rho _2^2+\tau _2^2+1\right)+m^2(\xi ^2 \rho _2^2+ \tau _2^2), \\
    gp = \frac{m^2 \left(1+u\right) \chi _l \left(\xi ^2 \rho _2^2+\tau _2^2+1\right)}{2 u \chi _g}+\frac{m^2 u \chi _g \left(\xi ^2 \rho _1^2+\tau _1^2+1\right)}{2 \left(1+u\right) \chi _l}\nonumber\\-m^2 (\xi ^2 \rho _1 \rho _2 + \tau _1 \tau _2), \\
    fp = \frac{m^2 \left(1+u\right) \chi _l \left(\xi ^2 \rho _2^2+\tau _2^2+1\right)}{2 u \chi _f}+\frac{m^2 u \chi _f \left(\xi ^2 \rho _1^2+\tau _1^2+1\right)}{2 \left(1+u\right) \chi _l}\nonumber\\+m^2 (\xi ^2 \rho _1 \rho _2+ \tau _1 \tau _2), \\
    gq = \frac{m^2 \chi _l \left(\xi ^2 \rho _2^2+\tau _2^2+1\right)}{2 u_1 \chi _g}+\frac{m^2 u_1 \chi _g \left(\xi ^2 \rho _1^2+\tau _1^2+1\right)}{2 \chi _l}-m^2( \xi ^2 \rho_1 \rho_2 + \tau_1 \tau_2), \\
    fq = \frac{m^2 \chi _l \left(\xi ^2 \rho _2^2+\tau _2^2+1\right)}{2 u_1 \chi _f}+\frac{m^2 u_1 \chi _f \left(\xi ^2 \rho _1^2+\tau _1^2+1\right)}{2 \chi _l}+m^2 (\xi ^2 \rho _1 \rho _2 + \tau _1 \tau _2).
\end{eqnarray}
The photon propagator in terms of the new variables is expressed as 
\begin{equation*}
    \frac{1}{(p-q+sk)^2+i\varepsilon} = \frac{\xi(1+u)}{2m^2u\chi} \times \frac{1}{s-\frac{u \xi}{2\chi}(1+\tau_1^2+\xi^2\rho_1^2)+i\varepsilon}.
\end{equation*}
A common phase factor arising from the vertices takes the form
$$\exp\big(i(\rho_1-\rho_2)(s-t)\big).$$
By a shift
\begin{align*}
    &s \rightarrow s + \frac{u \xi}{2\chi}(1+\tau_1^2+\xi^2\rho_1^2), \\
    &t \rightarrow t + \frac{u \xi}{2\chi}(1+\tau_1^2+\xi^2\rho_1^2),
\end{align*}
the dependence of the probability on $\rho_1$ and $\rho_2$ is reduced solely to this phase factor. 

For the sake of reliability, we use computer algebra for the above-described steps [namely, evaluation and contraction of traces, passing to variables eq.~\eqref{eq:new_vars} and shifting $s$ and $t$, see ref.~\cite{github}]. 
The resulting expression for the rate takes the form
\begin{equation}\label{eq:trident_prob_rate2}
    \begin{split}
        W_{\text{dir}} = \frac{1}{2} \frac{m^2 \alpha^2}{p^0} &\sum_{\sigma = \pm1}\int_0^\infty du\int_4^\infty dv \int_{-\infty}^{+\infty} d\tau_1\,d\tau_2 \\
        & \times \int_{-\infty}^{+\infty} \frac{d\rho_1\,d\rho_2}{L_\varphi} \int_{-\infty}^{+\infty} ds\,dt \,\frac{G(s,t) e^{i(\rho_1-\rho_2)(s-t)}}{(s+i\varepsilon)(t-i\varepsilon)},
    \end{split}
\end{equation}
\begin{align}\label{eq:G_func}
    &G(s,t) = \frac{2^{2/3} \xi  \left(u^2 (v-2)+2 (1+u) (v-4)\right)}{4\pi^2\, \chi^{1/3}\,u \,(1+u)^{7/3}\, v^{4/3} \,\sqrt{v(v-4)} }\,\text{Ai}'\left(y_{1s}\right) \text{Ai}'\left(y_{2s}\right) \text{Ai}'\left(y_{1t}\right) \text{Ai}'\left(y_{2t}\right) \nonumber \\
    & + \frac{\xi  \left(\tau _1^2 \left(u^2 (v-2)+2 (1+u) v\right)+u^2(v-2)\right)}{4\pi^2 \,\chi \, u^{1/3}\,(1+u)^{7/3} \, v^{4/3} \sqrt{v(v-4)}} \text{Ai}\left(y_{1s}\right) \text{Ai}'\left(y_{2s}\right) \text{Ai}\left(y_{1t}\right) \text{Ai}'\left(y_{2t}\right) \nonumber \\
    & + \frac{\xi  \left(\tau _2^2 \left(u^2 (v-2)+2 (1+u) v\right)+\left(u^2+2 u+2\right) v\right)}{4 \pi ^2  \chi u^{5/3} (1+u)^{5/3} v^{2/3} \sqrt{v(v-4)}}\text{Ai}'\left(y_{1s}\right) \text{Ai}\left(y_{2s}\right) \text{Ai}'\left(y_{1t}\right) \text{Ai}\left(y_{2t}\right) \nonumber \\
    & + \bigg[\xi^2 v \Big(\tau_1^2 \tau_2^2 \left(u^2 (v-2)+2 (1+u) (v-4)\right)+\tau_1^2\left(u^2+2 u+2\right) v + \tau_2^2 u^2(v-2) + u^2v\Big) \nonumber \\
    & \quad \quad + 4 \sigma \xi \tau_1 \tau_2 (u+2) \sqrt{v(v-4)} \chi (s+t)+32 s t \chi^2\bigg] \nonumber\\ &\quad \quad\times\frac{\text{Ai}\left(y_{1s}\right) \text{Ai}\left(y_{2s}\right) \text{Ai}\left(y_{1t}\right) \text{Ai}\left(y_{2t}\right)}{4\pi^2 2^{2/3}  \chi^{5/3} \xi u (1+u)^{5/3} v^{5/3} \sqrt{v(v-4)}} \nonumber \\
    & -\Bigg[\frac{\xi  \tau _1 \tau _2 v \left(u^2 (v-2)+2 (1+u) (v-4)\right) + 4 \sigma t \chi (u+2) \sqrt{v(v-4)} }{4 \pi ^2 \chi u (1+u)^2 v^2 \sqrt{v(v-4)}} \nonumber \\ 
    & \quad \quad \times \text{Ai}\left(y_{1s}\right) \text{Ai}\left(y_{2s}\right) \text{Ai}'\left(y_{1t}\right) \text{Ai}'\left(y_{2t}\right) \nonumber \\
    & \quad \quad +\frac{\xi  \tau_1 \tau_2 \left(u^2 (v-2)+2 (1+u) v\right)}{4\pi^2 \chi u (1+u)^2 v \sqrt{v(v-4)}} \text{Ai}'\left(y_{1s}\right) \text{Ai}\left(y_{2s}\right) \text{Ai}\left(y_{1t}\right) \text{Ai}'\left(y_{2t}\right) + (s \leftrightarrow t)\Bigg],
\end{align}
and $y_{1t}$, $y_{2t}$ differ from
\begin{equation}\label{eq:y_args}
\begin{split}
    &y_{1s} = \frac{s}{\xi}\left(\frac{2\chi}{u}\right)^{\frac{1}{3}} + \left(\frac{u}{2\chi}\right)^{\frac{2}{3}}(1+\tau_1^2), \\
    &y_{2s} = -\frac{s}{\xi}\left(\frac{2 u \chi}{v(1+u)}\right)^{\frac{1}{3}} + \left(\frac{v(1+u)}{2 u \chi}\right)^{\frac{2}{3}}(1+\tau_2^2).
\end{split}  
\end{equation}
by  replacing $s\to t$.

\subsubsection{Regularization of phase integrals}
Consider the inner integral from eq.~\eqref{eq:trident_prob_rate2}:
\begin{equation}
    \mathcal{I} = \int_{-\infty}^{+\infty} \frac{d\rho_1\,d\rho_2}{L_\varphi} \int_{-\infty}^{+\infty} ds\,dt \frac{G(s,t) e^{i(\rho_1-\rho_2)(s-t)}}{(s+i\varepsilon)(t-i\varepsilon)}.
\end{equation}
This expression is actually divergent and should be regularized. Since variables $\rho_1$ and $\rho_2$ are the formation phase instants of the two stages of the trident process, it is natural to bound their range to $\left[-\frac{L_\varphi}{2}, \frac{L_\varphi}{2}\right]$, where the already introduced $L_\varphi \gg 1$ is the phase extent of the field, see eq.~\eqref{eq:L_phi_def}. As a characteristics of the field, it is independent of any quantum numbers.
With that, we get the regularized expression
\begin{equation}
    \mathcal{I} = \int_{-L_\varphi/2}^{+L_\varphi/2} \frac{d\rho_1\,d\rho_2}{L_\varphi} \int_{-\infty}^{+\infty} ds\,dt \frac{G(s,t) e^{i(\rho_1-\rho_2)(s-t)}}{(s+i\varepsilon)(t-i\varepsilon)}.
\end{equation}
By performing the integration over $\rho_1$ and $\rho_2$, we arrive at
\begin{equation}
    \mathcal{I} = \frac{1}{L_\varphi}\int_{-\infty}^{+\infty} ds\, dt\, \frac{G(s,t)}{(s+i\varepsilon)(t-i\varepsilon)} \left(\frac{\sin\left(L_\varphi\frac{s-t}{2}\right)}{\frac{s-t}{2}}\right)^2.
\end{equation}
After passing to variables $$x = L_\varphi\frac{s-t}{2}, \quad T = \frac{s+t}{2},$$
it reads
\begin{equation}
    \mathcal{I} = 2 \int_{-\infty}^{+\infty} d T\, d x\, \frac{G(T+x/L_\varphi,\,T-x/L_\varphi)}{(T+x/L_\varphi+i\varepsilon)(T-x/L_\varphi-i\varepsilon)}\frac{\sin^2x}{x^2}.
\end{equation}
This integral is formed at $x \lesssim 1$. Assuming $L_\varphi \gg 1$, we expand the integrand as
\begin{equation}
    G(T+x/L_\varphi,\,T-x/L_\varphi) = G(T,T) + (G_s(T,T)-G_t(T,T)) \frac{x}{L_\varphi} + O\left(\frac{1}{L_\varphi^2}\right).
\end{equation}
Since $G(s,t)$ is symmetric with respect to interchanging the arguments, the second term in the expansion vanishes, and we discard all terms of the order $\left(1/L_\varphi\right)^2$ and higher. By adding and subtracting  $G(0,0)$ in the numerator, we separate the integral into two terms
\begin{align}
    &\mathcal{I} = \mathcal{I}_1 + \mathcal{I}_2, \\
    &\mathcal{I}_1 = 2 \int_{-\infty}^{+\infty} d T\, d x\, \frac{G(T,T) - G(0,0)}{(T+x/L_\varphi+i\varepsilon)(T-x/L_\varphi-i\varepsilon)}\frac{\sin^2x}{x^2}, \label{eq:one_step0} \\
    &\mathcal{I}_2 = 2 \int_{-\infty}^{+\infty} d T\, d x\, \frac{G(0,0)}{(T+x/L_\varphi+i\varepsilon)(T-x/L_\varphi-i\varepsilon)}\frac{\sin^2x}{x^2}. \label{eq:two_step0}
\end{align}
After dealing with integrations (see appendix~\ref{sec:app_a}), at the leading order in $1/L_\varphi$, we get
\begin{align}
    &\mathcal{I}_1 = 2\pi \int_0^\infty \frac{dT}{T^2}\left(G(T,T) + G(-T,-T) - 2G(0,0)\right), \label{eq:one_step1} \\
    &\mathcal{I}_2 = 2 \pi^2 L_\varphi G(0,0). \label{eq:two_step1}
\end{align}
Setting $s = t = 0$ corresponds to putting the intermediate photon on shell. Successive emission and pair production by an on-shell photon is conventionally referred to as the two-step process, and its rate is given by $\mathcal{I}_2$. On the other hand, $\mathcal{I}_1$ corresponds to the trident process mediated by an off-shell virtual photon, which is called the one-step process.

\subsubsection{Direct process rate}
Integrals over $\tau_1$ and $\tau_2$ are straightforward (see appendix~\ref{sec:app_b}). After integration and rescaling the variable $T$, we arrive at the following formula for the total rate
\begin{equation}\label{eq:rate_direct}
    \begin{split}
        W_{\text{dir}} = \frac{1}{2}\frac{m^2 \alpha^2}{p^0}\sum_{\sigma=\pm1}\int_0^\infty du & \int_4^\infty dv \Bigg[\xi L_\varphi \frac{\pi}{2}\left(\frac{u}{\chi}\right)^{\frac{1}{3}} \mathcal{G}(0) \\
         & + \int_0^\infty \frac{d\rho}{\rho^2}\left(\mathcal{G}(\rho) + \mathcal{G}(-\rho) - 2\mathcal{G}(0)\right) \Bigg],
    \end{split}
\end{equation}
where $\mathcal{G}(\rho) \equiv \mathcal{G}(\rho, u, v)$ is given by

\begin{align}\label{eq:G_cal_def}
    \mathcal{G}(\rho) =& \frac{4\pi}{\xi} \left(\frac{\chi}{u}\right)^{\frac{1}{3}} \int_{-\infty}^{+\infty}d\tau_1\,d\tau_2\,G\left(\rho \frac{\xi}{2}\left(\frac{u}{\chi}\right)^{\frac{1}{3}},\, \rho \frac{\xi}{2}\left(\frac{u}{\chi}\right)^{\frac{1}{3}}\right) =  \nonumber \\
     =& ~ \mathcal{C}^{++} \Ai'(z_1+\rho)\Ai'(z_2-\kappa\rho) + \mathcal{C}^{-+} \Ai_1(z_1+\rho)\Ai'(z_2-\kappa\rho) \nonumber \\
    & + \mathcal{C}^{+-} \Ai'(z_1+\rho)\Ai_1(z_2-\kappa\rho) + \mathcal{C}^{--} \Ai_1(z_1+\rho)\Ai_1(z_2-\kappa\rho) \nonumber \\
    & + \sigma \, \mathcal{C}^{00}\Ai(z_1+\rho)\Ai(z_2-\kappa\rho),
\end{align}
where 
\begin{align}
    &\label{eq:cpp}
    \mathcal{C}^{++} = \frac{u^2 (v-2)+(1+u) (2 v-5)}{\pi \, z_2 \, u^2 (1+u)^2 \, v \sqrt{v(v-4)}}, \\
    &\label{eq:cmp}
    \mathcal{C}^{-+} = \frac{(v-2) \left(\rho  \left(u^2+2 u+2\right)+2 (1+u) z_1\right)}{2\pi \, z_2 \, u^2 (1+u)^2 \, v \sqrt{v(v-4)}}, \\
    & \label{eq:cpm}
    \mathcal{C}^{+-} = -\frac{\left(u^2+2 u+2\right) \left(\rho  u^2 (v-2)+2 (1+u) v z_1\right)}{2\pi \,  z_1 \,  u^2 (1+u)^3 \, v^2 \sqrt{v(v-4)}}, \\
    &\mathcal{C}^{--} = -\big[\rho ^2 u^2 \left(u^2 (v-2)+2 (1+u) (v-6)\right)+4 \rho  (1+u) z_1 \left(u^2 (v-1)+(1+u) v\right) \nonumber \\
    & \label{eq:cmm}
    \quad \quad \quad \quad + 4 (1+u)^2 v z_1^2\big]\frac{1}{4\pi \, z_1 \, u^2 (1+u)^3 \, v^2 \sqrt{(v-4) v}}, \\
    \label{eq:c00}
    &\mathcal{C}^{00} = -\frac{\,\kappa \,\rho \, (u+2)}{\pi \, z_1\, u (u+1)^2\, v^2}.
\end{align}
and
\begin{equation}\label{eq:z1_z2_def}
    z_1 = \left(\frac{u}{\chi}\right)^{\frac{2}{3}}, \quad z_2 = \left(\frac{v(1+u)}{u \chi}\right)^{\frac{2}{3}}, \quad \kappa = \left(\frac{u^2}{v(1+u)}\right)^{\frac{1}{3}}.
\end{equation}

\subsection{Trident process rate from the two-loop scattering amplitude}
\label{sec:vi_2}
Now let us obtain the trident process rate from the two-loop scattering amplitude $\mT_{pp}^{(4),\mathrm{pol}}$ (recall the discussion in sections~\ref{sec:iv} and \ref{sec:v}).
According to eqs.~\eqref{eq:cut_trident}, \eqref{eq:cutting_equation_two_loop_explicit_amplitude}:\footnote{Recall that measure $d\Pi(q,f,g)$ in eq.~\eqref{eq:cut_trident} contains 1/2, see eq.~\eqref{eq:measure_Pi_tilde}.}
\[
\frac{1}{2p^0 VT} \times \includegraphics[align=c,scale=1]{massop_bubble/2loop_massop_a_trident_cut_lab.pdf}=W_{\mathrm{dir}}+W_{\mathrm{ex}}=2W_{\mathrm{dir}}
\]
where $W_{\rm dir,ex}$ denote the direct and exchange contributions, respectively.
To distinguish the result obtained in this section, we denote the corresponding rate by $\Wtr$. The LHS in this equation is obtained from $\mT_{pp}^{(4),\mathrm{pol}}=-VT\Im(\mM_0^{\rm (4,\,reg)}+\mM_1^{\rm (4)}+\mM_2^{\rm (4)})$ by using the substitution given in eq.~\eqref{substitution_trident2} in the expressions for $\mM_i^{(4)}$, see eq.~\eqref{M_two_loop}. This gives
\begin{equation}
    \label{Wtr_from_im_M}
		\begin{split}
			\Wtr(\chi)=&\frac{\alpha m^2\theta(p_-)}{2\pi p^0}\int_{0}^{\infty}\frac{du}{(1+u)^2}\, \int_{-\infty}^{\infty} d\lambda \left[-\left(\frac{\mP}{\lambda}\right)'+2\pi^2\delta^2(\lambda)\right] \\
            &\quad \times\left[\sum\limits_i\Im\Pi_i(\lambda,\chi_l)\,  \mathcal{R}_{i}(0,\lambda,u)-\left.\Im\Pi_0(\lambda,\chi_l)\,  \mathcal{R}_{0}(0,\lambda,u) \vphantom{\frac11}\right|_{F=0}\right].
		\end{split}
\end{equation}
Here, we used the equality $\theta(q_-)\theta(l_-)=\theta(p_-)\theta(u)$, given that $q_-=p_-/(1+u)$ and $l_-=u p_-/(1+u)$ [c.f. eq.~\eqref{eq:chi_ql}] and carried out the $\delta$-functional integral over $\mu$. As we mentioned, we subtract $\mM_i^{(4)}$ at zero field to regularize the vacuum divergence. Note that eq.~\eqref{Wtr_from_im_M} corresponds to the trident process initiated by a \textit{spin-polarized} electron. In this calculation, we leave the corresponding spin averaging over for the last step. 

To proceed, we plug $\Im \Pi_i$ into eq.~\eqref{Wtr_from_im_M}, where, as follows from eqs.~\eqref{pi0}, \eqref{pi12},
\begin{equation}\label{im_pi0}
		\Im\Pi_0(\lambda,\chi_l)=-4\alpha \lambda  \int_4^\infty \frac{dv}{v^{5/2}\sqrt{v-4}} \Ai_1(\zeta),
\end{equation}
\begin{equation}\label{im_pi12}
    \begin{split}
	\Im\Pi_{1,2}(\lambda,\chi_l)=-2\alpha m^2\int_4^\infty \frac{dv}{v^{3/2}\sqrt{v-4}} &\left[\left(1+\lambda\frac{v-6}{2v}\right)\Ai_1(\zeta)\right.\\
    &\quad\left.-(v-2\mp1)\left(\frac{\chi_l}{v}\right)^{2/3}\Ai'(\zeta)\right].
    \end{split}
\end{equation}
In the first expression, we used the equality $\Ai_1^{\rm (reg)}(\zeta)+\theta(-\zeta)=\Ai_1(\zeta)$. Furthermore, we rewrote $\Im\Pi_{1,2}$ by integrating eq.~\eqref{pi12} by parts. The integrand in this expression at $\lambda=0$ recovers the proper differential pair photoproduction rate \cite{ritus1985quantum,king2013photon}. The two terms in the first square bracket of eq.~\eqref{Wtr_from_im_M} correspond to one- and two-step terms, $\Wtr=\Wtr^{\rm 1\,st}+\Wtr^{\rm 2\,st}$. Let us consider them one by one.

\subsubsection{One-step contribution}
\label{sec:vi_2_1}
Consider the term in eq.~\eqref{Wtr_from_im_M} that is proportional to $\psi(\lambda)=-\left(\frac{\mP}{\lambda}\right)'$. The action of the distribution $\psi(\lambda)$ on a test function $\phi(\lambda)$ can be represented as a finite integral:\footnote{To show this, we first transfer in pairing the distributional derivative from $\psi$ to $\phi$ to obtain the integral prescribed by the principal value. Then we take it by parts to remove the derivative from $\phi'(\lambda)$, picking up $\phi(\lambda)-\phi(0)$ as its antiderivative. Next, we redefine the integration variable $\lambda$ in the integral over negative $\lambda$ by 
flipping its sign and combining it back with the integral over positive $\lambda$. Finally, by noticing that the resulting integrand is nonsingular at zero, we can abandon the principal value prescription.}
\begin{equation}
    \left\langle \psi,\,\phi\right\rangle = \int_0^\infty d\lambda\frac{\phi(\lambda)+\phi(-\lambda)-2\phi(0)}{\lambda^2}.
\end{equation}
After substituting eqs.~\eqref{im_pi0}, \eqref{im_pi12} into eq.~\eqref{Wtr_from_im_M} and further algebraic simplifications, we arrive at the one-step contribution:
\begin{equation}
    \label{eq:W_trident_from_loops_1st}
    \begin{split}
    \Wtr^{\rm 1\,st}(\chi)=\frac{\alpha^2 m^4}{\pi p^0 } \theta(\chi) \int_{0}^{\infty}  du \int_4^\infty dv \int_0^\infty \frac{d\lambda}{\lambda^2} \frac{\phi(\lambda)+\phi(-\lambda)-2\phi(0)}{{(1+u)^2v^{3/2}\sqrt{v-4}}},
    \end{split}
\end{equation}
where
\begin{equation}
    \label{eq:phi}
    \begin{split}
        \phi(\lambda)=&\chi^{4/3}\frac{u^2(v-2)+(1+u)(2v-5)}{(1+u)^{5/3}v^{2/3}} \Ai'(t)\Ai'(\zeta) \\
       & + \left[\frac{\chi u}{(1+u)v}\right]^{2/3}\left(1+\frac{\lambda}{m^2}\frac{u^2+2u+2}{2u^2}\right)(v-2)\Ai_1(t)\Ai'(\zeta) \\
       & - \left(\frac{\chi}{u}\right)^{2/3}\frac{u^2+2u+2}{1+u}\left(1+\frac{\lambda}{m^2}\frac{v-2}{2v}\right)\Ai'(t)\Ai_1(\zeta)\\
       &-\left[1+\frac{\lambda}{m^2}\left(\frac{u^2+u+1}{u^2}-\frac{1}{v}\right)\right.\\
       &\quad\quad\,\,\left.+\frac{\lambda^2}{m^4}\frac{u^2(v-2)+2(1+u)(v-6)}{4u^2v}\right]\Ai_1(t)\Ai_1(\zeta)\\
      & + 2\gamma_s\frac{u^{4/3}(u-v+3)}{(1+u)^{5/3}v^{2/3}}\Ai(t)\Ai'(\zeta)\\
      & +  \frac{2\gamma_s}{1+u}\left(\frac{u}{\chi}\right)^{2/3}\left(1+\frac{\lambda}{m^2}\frac{v-2}{2v}\right)\Ai(t)\Ai_1(\zeta).
    \end{split}
\end{equation}
Here, $t$ and $\zeta$ are given by eqs.~\eqref{t} (with $\mu=0$) and \eqref{zeta}, respectively. Averaging over the initial electron spin projection, in effect, removes the last terms proportional to $\gamma_s$, and the rest of the terms amount to the rate for the process initiated by an unpolarized electron.

\subsubsection{Two-step contribution}
\label{sec:vi_2_2}
Now consider the term proportional to $\delta^2(\lambda)=\delta(\lambda=0)\delta(\lambda)$ in eq.~\eqref{Wtr_from_im_M}. Following refs.~\cite{ritus1972vacuum, king2013trident}, we associate it with the two-step process $\Wtr^{\mathrm{2\,st}}$, in which the intermediate photon is on-shell. 
The infinite prefactor $\delta(\lambda=0)$ is interpreted as a proportionality to a large time $T_\gamma$, or, alternatively, to a large phase (dimensionless light-front time) interval $L_\varphi$, during which the intermediate photon can travel in between the events of emission and pair creation. As the field is assumed constant, it lasts long and occupies a large volume, so that these points can be separated by a large distance. This is conventionally expressed as follows [c.f. eq.~\eqref{eq:L_phi_def}]:
\begin{equation}\label{eq:delta_zero}
    \begin{split}
    \delta(\lambda=0)=\frac{1}{2l_-}\delta(l_+-\omega_l=0)\simeq \frac{1}{4\pi \omega l_-}&\left.\int_{-L_\varphi/2}^{L_\varphi/2} d\varphi\,  e^{i\varphi(l_+-\omega_l)/\omega}\right|_{l_+=\omega_l}\\
        &\quad\quad\quad\quad\quad=\frac{L_\varphi}{4\pi (kl)}=\frac{\xi L_\varphi}{4\pi m^2 \chi}\frac{1+u}{u},
    \end{split}
\end{equation}
where we assume that $l_->0$, which allows us to pass to light-front coordinates in the first step, $\omega_l=l_\perp^2/2l_-$, and, in the second step, we regularised the delta function representation by introducing large finite limits (recall that $\omega=k^0$).

Integration over $\lambda$ for the two-step term is straightforward. Note that, since $\Pi_0(\lambda,\chi_l)\propto\lambda$, the $\mM_0^{\rm (2,reg)}$ term does not contribute to the two-step rate. To obtain the final result, we substitute $\Im\Pi_{1,2}$ from eq.~\eqref{im_pi12} into eq.~\eqref{Wtr_from_im_M} and combine the contributions $\mM_{1,2}$. Finally, the two-step rate acquires the form:
\begin{equation}
    \label{eq:W_trident_from_loops_2st}
    \begin{split}
    \Wtr^{\mathrm{2\,st}}(\chi)=\frac{\alpha^2 m^2\theta(\chi)}{2 p^0 \chi} & \xi L_\varphi   \int_{0}^{\infty}  du \int_4^\infty dv \frac{\phi(0)}{u(1+u)v^{3/2}\sqrt{v-4}},
    \end{split}
\end{equation}
where $\phi(0)$ is derived from eq.~\eqref{eq:phi}. As for eq.~\eqref{eq:W_trident_from_loops_1st}, the spin-averaged rate is obtained by omitting the last two terms proportional to $\gamma_s$.

\subsubsection{Comparison of the loop and tree-level calculations}
\label{sec:vi_2_3}
Finally, let us compare rate $W_{\text{dir}}$, obtained from the scattering amplitude defined at the tree level [see eqs.~\eqref{eq:rate_direct}--\eqref{eq:cmm}], to rate $\Wtr$ resulting from cutting the two-loop diagram [see eqs.~\eqref{eq:W_trident_from_loops_1st}, \eqref{eq:phi}, \eqref{eq:W_trident_from_loops_2st}]. For this, let us rewrite the one step contribution $W_{\text{dir}}^{\mathrm{1\,st}}$ in the notations of $\Wtr^{\mathrm{1\,st}}$. This is achieved by changing variable $\rho$ to $\lambda$ as $\rho=(\lambda/m^2)(1+u)/u^{4/3}\chi^{2/3}$. Then, after summing over $\sigma$ [see \eqref{eq:rate_direct}], we recover the one-step contribution obtained by averaging eq.~\eqref{eq:W_trident_from_loops_2st} over the initial electron spin projection (namely, by setting $\gamma_s\rightarrow0$). It is also straightforward to see that the two-step contribution to eqs.~\eqref{eq:rate_direct}--\eqref{eq:cmm} matches eq.~\eqref{eq:W_trident_from_loops_2st} after the change of variables $\rho\rightarrow\lambda$ and spin averaging. Hence, the result of cutting exactly matches the tree-level calculation. 

Let us add two comments. First, both integrands entering the rates given in eqs.~\eqref{eq:rate_direct}--\eqref{eq:cmm} and eqs.~\eqref{eq:W_trident_from_loops_1st}, \eqref{eq:phi}, and \eqref{eq:W_trident_from_loops_2st} provide proper differential distributions in quantum parameters $\chi_{q,f,g}$ of the final particles. We rewrite these distributions in more natural variables and analyse them below in Section~\ref{sec:vii}. Second, we observe that the approach relying on cutting the two-loop electron elastic scattering amplitude does not guarantee that the result will have the shape of the proper differential rate, integrated over the dynamical variables. Likewise, it is unlikely that the result reported by Ritus in ref.~\cite{ritus1972vacuum} has such a form. The calculation presented therein relies on $\Im\Pi_i$ as given in eqs.~\eqref{pi0}, \eqref{pi12}. While $\Im\Pi_i$ in such a form still links to the total probability rate of pair creation by a polarized photon, the integrand is not shaped as a proper differential distribution \cite{ritus1985quantum}. This means that information about the pair spectrum is distorted from the start, and hence cannot be seen in the final expressions for the trident rate as derived in  ref.~\cite{ritus1972vacuum}. It is worth noting, though, that it is never stated in ref.~\cite{ritus1972vacuum} that the formula for the trident rate derived therein contains a true differential distribution.

\section{Distributions of the quantum parameters in  trident pair production}
\label{sec:vii}
In this section, we discuss distributions of the quantum parameters of the outgoing electrons in the trident process. Recall the tree-level result \eqref{eq:rate_direct}--\eqref{eq:cmm} for the direct contribution to the trident pair production rate. For convenience, we introduce new variables 
\begin{equation}
    k = \frac{\chi_q}{\chi} = \frac{1}{1+u}, \quad r = \frac{\chi_g}{\chi} = \frac{u}{2(1+u)}\left(1-\sigma\sqrt{1-\frac{4}{v}}\right).
\end{equation}
The inverse relations and the Jacobian are given by
\begin{equation}
    u = \frac{1-k}{k}, \quad v = \frac{(1-k)^2}{r(1-k-r)}, \quad \left|\frac{\partial(u,v)}{\partial(k,r)}\right| = \frac{(1-k)^2|k+2r-1|}{k^2\,r^2\,(1-k-r)^2}.
\end{equation}
Note that $\sigma = +1$ corresponds to the region $r \leq (1-k)/2$ and $\sigma = -1$ to the region $r \geq (1-k)/2$. Next, we define the new auxiliary function
\begin{align}
    \mathcal{Q}(\rho; k, r) =& \frac{(1-k)^2|k+2r-1|}{k^2\,r^2\,(1-k-r)^2} \mathcal{G}\left(\rho;\, u = \frac{1-k}{k},\, v = \frac{(1-k)^2}{r(1-k-r)}, \,\sigma = \operatorname{sgn}(1-k-2r)\right) \nonumber \\
    =& ~\mathcal{K}^{++} \Ai'(z_1+\rho)\Ai'(z_2-\kappa\rho) + \mathcal{K}^{-+} \Ai_1(z_1+\rho)\Ai'(z_2-\kappa\rho) \nonumber \\
    & + \mathcal{K}^{+-} \Ai'(z_1+\rho)\Ai_1(z_2-\kappa\rho) + \mathcal{K}^{--} \Ai_1(z_1+\rho)\Ai_1(z_2-\kappa\rho) \nonumber \\
    & + \mathcal{K}^{00}\Ai(z_1+\rho)\Ai(z_2-\kappa\rho),
\end{align}
where 
\begin{align}
    \mathcal{K}^{++} =& \frac{k^4+2 (k^3+r) (r-1)+k^2 \left(2 r^2-r+2\right)+k \left(r^2+r-2\right)+1}{\pi \, z_2 \, (1-k)^3 \,r\, (1-k-r)}, \\
    \mathcal{K}^{-+} =& \frac{\left(k^2+2 (k+r) (r-1)+1\right) \left((k^2+1)\rho +2 k z_1 \right)}{2 \pi \, z_2 \, (1-k)^3 \,r \,(1-k-r)}, \\
    \mathcal{K}^{+-} =& -\frac{\left(k^2+1\right) \left(\rho  \left(k^2+2 (k+r) (r-1)+1\right)+2 k z_1\right)}{2 \pi \, z_1 \, (1-k)^3 \,k}, \\
    \mathcal{K}^{--} =& -\Big[\rho ^2 \left(k^4+2 (k^3+r) (r-1)+2 k^2 \left(r^2+3 r+1\right)+k \left(8 r^2-6 r-2\right)+1\right) \nonumber\\
    & + 4 k \rho  z_1 \left(k^2+(k+r) (r-1)+1\right)+4 k^2 z_1^2\Big]\frac{1}{4 \pi \, z_1 \,(1-k)^3\, k}. \\
    \mathcal{K}^{00} =& \frac{\kappa \rho (k+1)(k+2 r-1)}{\pi \,z_1\, (1-k)^3}
\end{align}
and 
\begin{equation}
    z_1 = \left(\frac{1-k}{k \chi}\right)^{\frac{2}{3}}, \quad z_2 = \left(\frac{1-k}{r(1-k-r)\chi}\right)^{\frac{2}{3}}, \quad \kappa = \left(\frac{r(1-k-r)}{k}\right)^{\frac{1}{3}}.
\end{equation}
Then we can rewrite the expression for the trident rate in the form
\begin{align}\label{eq:Wdir_kr}
    W_{\text{dir}} = \frac{1}{2}\frac{m^2 \alpha^2}{p^0}\int_0^1dk\int_0^{1-k}dr \, w_{\text{dir}}(k,r),
\end{align}
where
\begin{equation}\label{eq:Wdir_kr_diff}
    \begin{split}
        & w_{\text{dir}}(k,r) = w_{\text{dir}}^{\mathrm{2\,st}}(k,r) + w_{\text{dir}}^{\mathrm{1\,st}}(k,r), \\
        & w_{\text{dir}}^{\mathrm{2\,st}}(k,r) = \xi L_\varphi \frac{\pi}{2}\left(\frac{1-k}{k \chi}\right)^{\frac{1}{3}} \mathcal{Q}(0; k, r), \\
        & w_{\text{dir}}^{\mathrm{1\,st}}(k,r) = \int_0^\infty \frac{d\rho}{\rho^2}\left(\mathcal{Q}(\rho; k, r) + \mathcal{Q}(-\rho; k, r) - 2\mathcal{Q}(0; k, r)\right).
    \end{split}
\end{equation}
The function $w_{\text{dir}}(k,r)$ gives the direct contribution to the distribution density in the $(k,r)$ plane in units of the characteristic rate $m^2 \alpha^2/p^0$. To obtain the total rate, we should add the exchange term. Since the direct and exchange diagrams differ only by an exchange of electrons, it follows that the distribution density for the exchange term is
\begin{equation}
    w_{\text{ex.}}(k, r) = w_{\text{dir}}(r,k)
\end{equation}
and we arrive at
\begin{align}
    &\frac{d^2 W_{\text{no-int.}}}{dk \, dr} = \frac{m^2\alpha^2}{p^0} (w_{\text{dir}}(k,r) + w_{\text{dir}}(r,k)) \equiv \frac{m^2\alpha^2}{p^0} ( w_{\mathrm{2\,st}}(k,r) + w_{\mathrm{1\,st}}(k,r)),\\
    & w_{\mathrm{2\,st}}(k,r) = \xi L_\varphi \frac{\pi}{2} \left(\left(\frac{1-k}{k \chi}\right)^{\frac{1}{3}} \mathcal{Q}(0; k, r) + \left(\frac{1-r}{r \chi}\right)^{\frac{1}{3}} \mathcal{Q}(0; r, k)\right), \\
    & w_{\mathrm{1\,st}}(k,r) = \int_0^\infty \frac{d\rho}{\rho^2}\left(\mathcal{Q}(\rho; k, r) + \mathcal{Q}(-\rho; k, r) - 2\mathcal{Q}(0; k, r)\right) + (k \leftrightarrow r),
\end{align}
where $W_{\text{no-int.}}$ is the total rate for the trident process excluding the interference term. Functions $w_{\mathrm{2\,st}}(k,r)$ and $w_{\mathrm{1\,st}}(k,r)$ are the two-step and one-step contributions to the total distribution density, respectively.

\begin{figure}
    \centering
    \scalebox{0.76}{\includegraphics[width=1\linewidth]{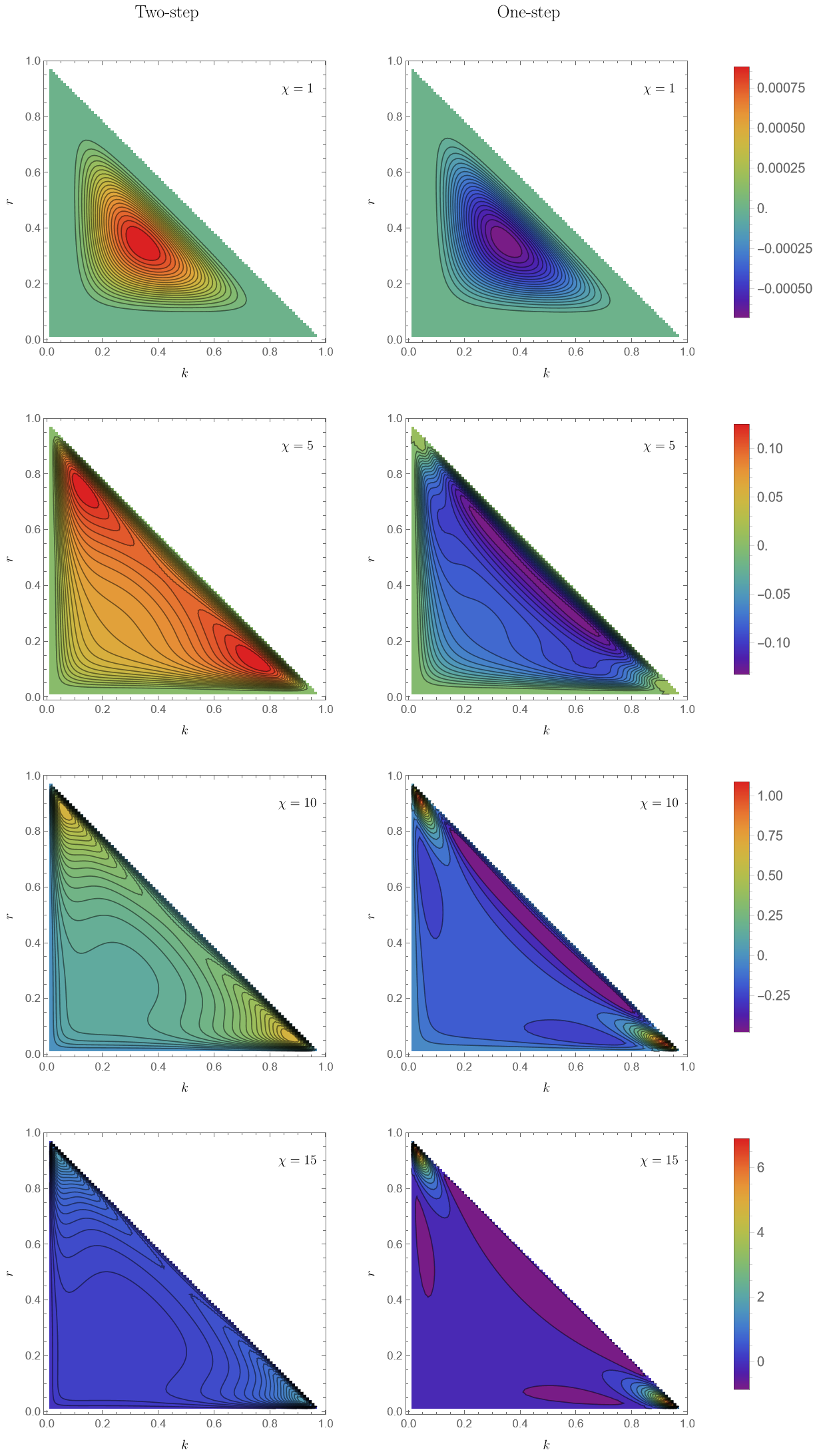}}
    \caption{Plots of differential rates of the two-step process $w_{\mathrm{2\,st}}(k,r)$ for $\xi L_{\varphi}=1$ (left column) versus differential rates of the one-step process $w_{\mathrm{1\,st}}(k,r)$ (right column) for different values of $\chi$.}
    \label{fig:distributions}
\end{figure}

Figure \ref{fig:distributions} shows the functions $w_{\mathrm{2\,st}}$ and $w_{\mathrm{1\,st}}$ for different values of the parameter $\chi$. As $\chi$ increases, the single maximum of $w_{\mathrm{2\,st}}$ splits into two peaks that migrate toward the points $k=1$ and $r=1$, becoming narrower and larger in magnitude. The one-step distribution $w_{\mathrm{1\,st}}$ behaves more trickily: at relatively small $\chi$, it remains negative over the entire $(k,r)$ domain, and with growing $\chi$, its sole minimum shifts to the line $k+r=1$, spreading along this line. However, above a certain $\chi$ threshold, positive maxima arise near $k=1$ and $r=1$. These maxima are similar to those of the two-step distribution $w_{\mathrm{2\,st}}$ but are narrower and attain higher values. The emergence of these peaks, in turn, induces the formation of additional minima.

\begin{figure}
    \centering
    \includegraphics[width=\linewidth]{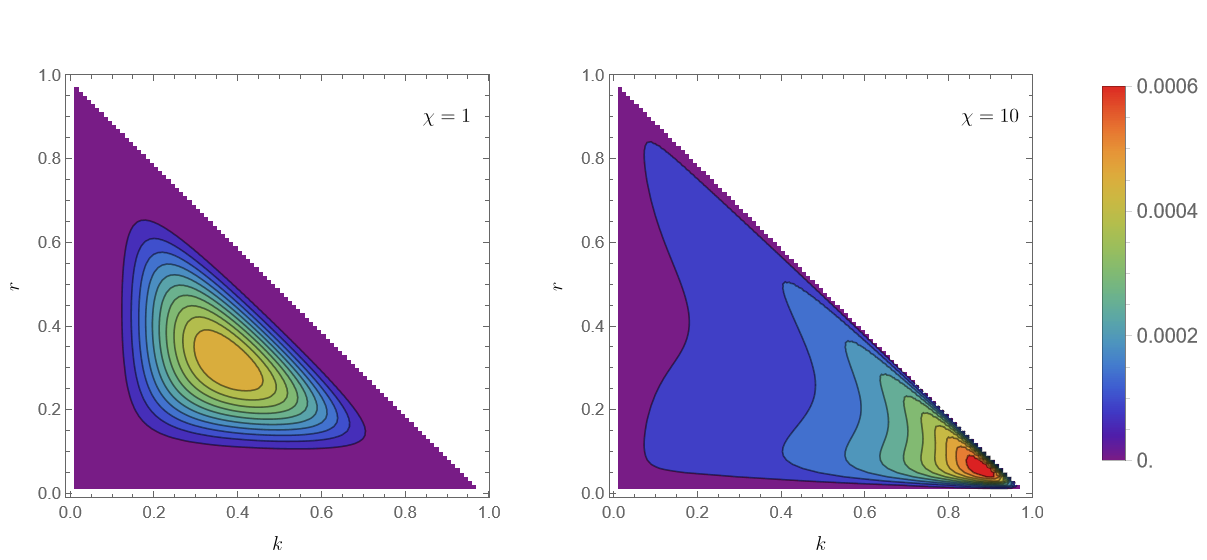}
    \caption{The differential rate of the direct two-step contribution $(1/\chi^3)w_{\text{dir}}^{\mathrm{2\,st}}(k,r)$ for $\chi=1$ and $\chi=10$.}
    \label{fig:King_two-step}
\end{figure}

\begin{figure}
    \centering
    \includegraphics[width=\linewidth]{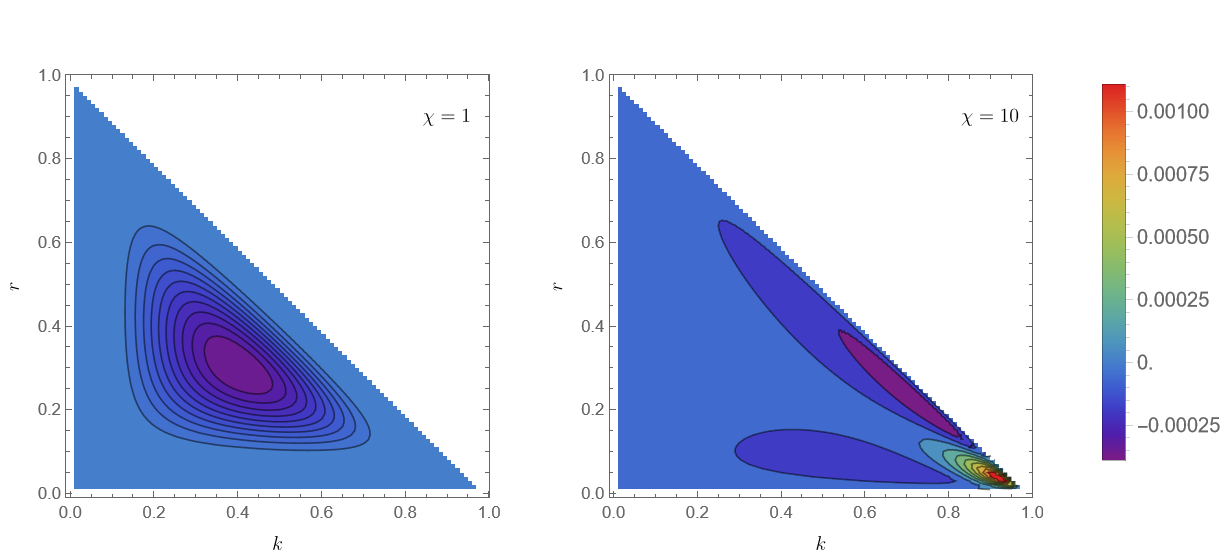}
    \caption{The differential rate of the direct one-step contribution $(1/\chi^3)w_{\text{dir}}^{\mathrm{1\,st}}(k,r)$ for $\chi=1$ and $\chi=10$.}
    \label{fig:King_one-step}
\end{figure}

For verification, the direct two- and one-step contributions were plotted separately for $\chi = 1$ and $\chi = 10$ in figure \ref{fig:King_two-step} and figure \ref{fig:King_one-step}, respectively, for convenience with the scaling factor $1/\chi^3$. The resulting distributions up to a proportionality coefficient agree with those previously reported in the literature \cite{king2013trident}.

\clearpage

\section{Conclusion}
\label{sec:viii}
In this work, we addressed the general problem of formulating cutting rules in QED in a strong plane-wave background. Although some applications of the optical theorem for computing process rates are known at one- and two-loop order~\cite{ritus1972radiative,ritus1972vacuum,tsai1974photon,meuren2015polarization,baier1972higher,baier1976theory,morozov1975elastic,morozov1977elastic}, the systematic formulation of such rules has generally been overlooked. To fill this gap, we derive the general version of the cutting equation for QED in a plane-wave background field using Veltman's approach~\cite{veltman1994diagrammatica} and identify the corresponding cutting rules. These rules turn out to be identical to those introduced on an ad hoc basis in refs.~\cite{ritus1972vacuum,morozov1975elastic,morozov1977elastic} for scattering processes in a CCF.

In ordinary QED, unitary cuts can be formulated as relations for invariant amplitudes, as it is possible to factor 4-momentum-conserving delta-functions. In the presence of a background, the latter does not hold, and hence scattering matrix elements [given by integrals over partial amplitudes, see eq.~\eqref{eq:s_matrix_sfqed}] should be considered instead. As an illustration, we considered the cutting of the electron forward scattering matrix element $\mT_{pp}$. Initially, the optical theorem is formulated for the total matrix element at a given order of perturbation theory. However, a stronger version can be formulated for individual gauge-invariant diagrams or for sums of all gauge-dependent diagrams at a given order — this is the cutting equation. We illustrate this by considering the two-loop $\mT^{(4)}_{pp}$ in a plane wave, as shown in eq.~\eqref{eq:T4_diagrams}. The cutting equation demonstrates that the optical theorem indeed applies to the total matrix element $\mT^{(4)}_{pp}$ [see eqs.~\eqref{eq:cutting_equation_two_loop_diagrams_2} and~\eqref{eq:optical_theorem_2_loop}]. For an explicit verification of the cutting equation, we provide a detailed reconsideration of the polarization correction $\mT^{(4),\mathrm{pol}}_{pp}$ [the first diagram in eq.~\eqref{eq:T4_diagrams}] in a CCF.

The expression for $\mT^{(4),\mathrm{pol}}_{pp}$ in a CCF was originally derived by Ritus~\cite{ritus1972vacuum}. We reproduce it in a different notation starting from the all-order resummed bubble-chain amplitude~\cite{mironov2020resummation}. In both approaches, most momentum integrals are already evaluated at the amplitude level. Nevertheless, it is still possible to apply unitary cuts in $p$-space, as formulated in section~\ref{sec:iii}. According to the cutting equation, this diagram decomposes into contributions corresponding to the trident process (the direct contribution) and to the correction to photon emission, as given by eqs.~\eqref{eq:cut_trident} and~\eqref{eq:cut_rad_corr}. Moreover, an explicit calculation (see section~\ref{sec:v}) demonstrates that summing these contributions indeed reproduces the imaginary part of $\mT^{(4),\mathrm{pol}}_{pp}$.

One nontrivial result emerging from the cutting equation for $\mT^{(4)}_{pp}$ (and specifically for $\mT^{(4),\mathrm{pol}}_{pp}$) is the appearance of field-induced, UV-finite loop corrections to outgoing particle states, for instance, the polarization correction to an outgoing photon line. Furthermore, these contributions help to cancel ambiguities that appear term by term on the right-hand side of the cutting equation. To see this, recall the two cuttings of $\mT^{(4),\mathrm{pol}}_{pp}$ that yield eqs.~\eqref{substitution_trident} (trident) and~\eqref{substitution_rad_pol_initial} (the polarization radiation correction). Both contain the same formally infinite term with opposite signs, which is attributed to the two-step trident pair production. When the two cuttings are summed, the infinite terms cancel, yielding a regular expression for $\Im \mT^{(4),\mathrm{pol}}_{pp}$.

This observation can be reformulated as follows. The rate (or cross-section) of a process in a background field can be extracted from a higher-order diagram only by applying appropriate cutting rules, since they allow one to reconstruct multi-step contributions. The imaginary part alone does not carry complete information about these contributions because they cancel exactly in the sum over all possible cuts on the right-hand side of the cutting equation.

Accordingly, using the cutting rules, we computed the direct contribution to the trident process rate, including both the one-step term [summarized in eqs.~\eqref{eq:W_trident_from_loops_1st},~\eqref{eq:phi}] and the two-step term [eq.~\eqref{eq:W_trident_from_loops_2st}]. Notably, this rate is resolved with respect to the spin states of the initial electron.

As a fully consistent check, we calculated the tree-level rate of the trident process in a CCF (the direct term) as defined by the matrix element. Since this calculation is lengthy, we outline the key steps of the derivation in section~\ref{sec:vi} and provide a full calculation in a computer algebra script available on \texttt{GitHub} \cite{github}. Although analogous calculations have been presented in earlier works~\cite{king2013trident,mackenroth2018nonlinear,dinu2018trident,torgrimsson2020nonlinear}, we obtain a new compact expression for the one-step contribution; see eqs.~\eqref{eq:rate_direct}--\eqref{eq:c00}. Within this calculation, we introduce a consistent regularization procedure for the divergent phase integrals. In the original work~\cite{ritus1972vacuum}, Ritus relies on the ill-defined squared $\delta$-function. In contrast, our approach employs a physically reasonable regulator, namely the phase extent of the field, and differs from that presented in~\cite{king2013trident}. Furthermore, we provide the differential probability rates [see eqs.~\eqref{eq:Wdir_kr},~\eqref{eq:Wdir_kr_diff}] expressed in units of the outgoing particle quantum parameters relative to the initial electron $\chi$.

The direct comparison of the trident rate extracted from the tree-level computation and the loop result for $T^{(4),\mathrm{pol}}_{pp}$ shows that, upon averaging over the initial electron spin, the expressions obtained within these approaches coincide [namely, our resulting eqs.~\eqref{eq:rate_direct}--\eqref{eq:c00} for the former match
eqs.~\eqref{eq:W_trident_from_loops_1st},~\eqref{eq:phi}, and~\eqref{eq:W_trident_from_loops_2st} for the latter]. This is one of our main results, illustrating the practical side of using unitary cuts and the related nuances arising during the calculation. 

We emphasize that these results match at the level of the \textit{the differential distribution rates}. However, the cutting equations do not guarantee this per se, as generally they connect expressions for \textit{total rates}. In our case, a stronger result became possible because, when cutting $T^{(4),\mathrm{pol}}_{pp}$, we used eqs.~\eqref{im_pi0},~\eqref{im_pi12} for the polarization operator imaginary part, which recover the correct energy distribution for the nonlinear Breit--Wheeler pair production when put on the mass shell. An alternative representation for $\Pi_{\mu\nu}$, e.g. given in eqs.~\eqref{pi0},~\eqref{pi12}, would yield an equivalent total trident process rate, but not the correct differential rate, when used in the cutting method. 

The proposed cutting rules can be applied at higher loop orders. However, in the present work, we have not discussed the aspects of renormalization at high orders of perturbation theory. A specific issue here is the identification of the particle mass shell for states that include loop corrections. In SFQED, such corrections are generally complex (e.g., the polarization operator) and therefore lead to the appearance of unstable states in the spectrum. This will be addressed in subsequent studies. As a possible generalization for future work, we hope that Veltman's approach could be employed to define unitarity-preserving cutting rules for QED in a strong CCF in the fully nonperturbative regime, namely for diagrams with all-order bubble-chain photon propagators~\cite{mironov2020resummation,mironov2022structure}.

\section*{Acknowledgments}
We are grateful to A. Ilderton, T. Adamo, F. Karbstein, B. King, H. Gies, and D. Seipt for fruitful discussions.
Y.V.S., A.I.A. and A.M.F. acknowledge support from the Foundation for the Advancement of
Theoretical Physics and Mathematics “BASIS” (Grant
No. 24-1-1-21).

\appendix
\section{Integrals encountered in the trident process separation procedure}
\label{sec:app_a}
First, we evaluate the integral \eqref{eq:two_step0} for the two-step process
\begin{equation}
    \mathcal{I}_2 = 2 \int_{-\infty}^{+\infty} d T\, d x\, \frac{G(0,0)}{(T+x/L_\varphi+i\varepsilon)(T-x/L_\varphi-i\varepsilon)}\frac{\sin^2x}{x^2}.
\end{equation}
The integral over $T$ is straightforward
\begin{align}
     &\int_{-\infty}^{+\infty} \frac{d T}{(T+x/L_\varphi+i\varepsilon)(T-x/L_\varphi-i\varepsilon)}  = \frac{i\pi L_\varphi}{x + i\varepsilon},
\end{align}
where we have redefined $\varepsilon L_\varphi \rightarrow \varepsilon$. The remaining integral over $x$ is of the form
\begin{align}
    &\mathcal{I}_2 = 2\pi i L_\varphi \int_{-\infty}^{+\infty}d x \frac{\sin^2 x}{x^2(x+i\varepsilon)} \equiv 2\pi i L_\varphi J, \nonumber \\
    &J =-i\varepsilon \int_{-\infty}^{+\infty} d x \frac{\sin^2 x}{x^2(x^2+\varepsilon^2)}.
\end{align}
Next, the integral is split into two parts
\begin{equation}
    J = -\frac{i}{\varepsilon}\int_{-\infty}^{+\infty} d x \sin^2 x \left(\frac{1}{x^2}-\frac{1}{x^2+\varepsilon^2}\right) = -\frac{i}{\varepsilon}\int_{-\infty}^{+\infty} d x\frac{\sin^2 x}{x^2} +\frac{i}{\varepsilon}\int_{-\infty}^{+\infty} d x\frac{\sin^2 x}{x^2+\varepsilon^2}.
\end{equation}
The first part is turned into the Dirichlet integral
\begin{equation}
    \int_{-\infty}^{+\infty} d x\frac{\sin^2x}{x^2} = \pi.
\end{equation}
Since the integrand in the second part no longer has a pole at $x = 0$, we can express $\sin^2 x$ as $(1-\cos(2x))/2$ and then use Jordan's lemma together with the residue theorem to evaluate the integral
\begin{equation}
    \frac{1}{2}\int_{-\infty}^{+\infty} \frac{d x}{x^2 + \varepsilon^2} - \frac{1}{2}\operatorname{Re} 
    \int_{-\infty}^{+\infty} \frac{d x e^{2ix}}{x^2+\varepsilon^2} = \frac{\pi}{2\varepsilon}\left(1-e^{-2\varepsilon}\right).
\end{equation}
Therefore, we arrive at
\begin{equation}
    J = -\frac{i\pi}{\varepsilon}\left(1 - \frac{1}{2\varepsilon}\left(1-e^{-2\varepsilon}\right)\right) = -i\pi + O(\varepsilon), ~\varepsilon \ll 1.
\end{equation}
Combining everything, the two-step integral is
\begin{equation}
    \mathcal{I}_2 = 2 \pi^2 L_\varphi G(0,0).
\end{equation}

Next, consider the integral \eqref{eq:one_step0} for the one-step process
\begin{equation}
    \begin{split}
        \mathcal{I}_1 &= 2 \int_{-\infty}^{+\infty} d T\, d x\, \frac{G(T,T) - G(0,0)}{(T+x/L_\varphi+i\varepsilon)(T-x/L_\varphi-i\varepsilon)}\frac{\sin^2x}{x^2} \\
        & = 2 L_\varphi^2\int_{-\infty}^{+\infty} d T \big(G(T,T)-G(0,0)\big) I(L_\varphi T, L_\varphi\varepsilon),
    \end{split}
\end{equation}
where
\begin{equation}
    I(\zeta, \eta) = \int_{-\infty}^{+\infty} d x \frac{\sin^2 x}{x^2(\zeta +x+i\eta)(\zeta-x-i\eta)}, \quad \zeta>0, \quad \eta>0.
\end{equation}
Let us transform this expression
\begin{align}
    I(\zeta,\eta) =& \int_{-\infty}^{+\infty} d x \frac{\sin^2 x}{x^2} \frac{\zeta^2+\eta^2-x^2+2i\eta x}{\big((\zeta+x)^2+\eta^2\big)\big((\zeta-x)^2+\eta^2\big)} \nonumber \\
    =& \int_{-\infty}^{+\infty} d x \sin^2x \Bigg[\frac{1}{\zeta^2+\eta^2}\frac{1}{x^2} + \frac{\zeta^2-\eta^2}{2\zeta(\zeta^2+\eta^2)^2} \left(\frac{x}{(\zeta+x)^2+\eta^2}-\frac{x}{(\zeta-x)^2+\eta^2}\right) \nonumber \\ 
    & \quad \quad \quad \quad \quad \quad \quad +\frac{\zeta^2-3\eta^2}{2(\zeta^2+\eta^2)^2}\left(\frac{1}{(\zeta+x)^2+\eta^2}+\frac{1}{(\zeta-x)^2+\eta^2}\right)\Bigg] \nonumber \\
    =& \int_{-\infty}^{+\infty} d x \Bigg[ \frac{1}{\zeta^2+\eta^2} \frac{\sin^2 x}{x^2} + \frac{\zeta^2-\eta^2}{\zeta(\zeta^2+\eta^2)^2} \frac{x \sin^2 x}{(\zeta+x)^2+\eta^2} + \frac{\zeta^2-3\eta^2}{(\zeta^2+\eta^2)^2} \frac{\sin^2 x}{(\zeta+x)^2+\eta^2}\Bigg].
\end{align}
Now we can evaluate
\begin{align}
    &\int_{-\infty}^{+\infty} d x \frac{x \sin^2 x}{(\zeta+x)^2+\eta^2} = -\frac{\pi \zeta}{2\eta}\big(1-e^{-2\eta}\cos(2\zeta)\big)-\frac{\pi}{2}e^{-2\eta}\sin(2\zeta),
\end{align}
and
\begin{align}
    &\int_{-\infty}^{+\infty} d x \frac{\sin^2 x}{(\zeta+x)^2+\eta^2} = \frac{\pi}{2\eta}\left(1-e^{-2\eta}\cos(2\zeta)\right).
\end{align}
Combining together, we get
\begin{align}
    I(\zeta,\eta) &= \frac{\pi}{\zeta^2+\eta^2} - \frac{\pi \eta}{(\zeta^2+\eta^2)^2}\big(1-e^{-2\eta}\cos(2\zeta)\big) + \frac{\pi (\eta^2-\zeta^2)}{2\zeta(\zeta^2+\eta^2)^2}e^{-2\eta}\sin(2\zeta) \\
    &= \pi \frac{2\zeta-\sin(2\zeta)}{2\zeta^3} + O(\eta), ~ \eta \ll 1.
\end{align}
Substituting this into $\mathcal{I}_1$ we obtain
\begin{align}
    \mathcal{I}_1 = 2 \pi \int_{-\infty}^{+\infty} \frac{d T}{T^2} \big(G(T,T)-G(0,0)\big) \left(1 - \frac{\sin(2L_\varphi T)}{2L_\varphi T}\right).
\end{align}
Since only the even part of the integrand contributes, we can symmetrize 
\begin{align}
    \mathcal{I}_1 = 2 \pi \int_{0}^{\infty} \frac{d T}{T^2} \big(G(T,T)+G(-T,-T)-2G(0,0)\big) \left(1 - \frac{\sin(2L_\varphi T)}{2L_\varphi T}\right).
\end{align}
For the last term in the parenthesis we use the representation of the $\delta$-function
\begin{equation}
    \underset{a\rightarrow\infty}{\operatorname{lim}}\frac{\sin(ax)}{x} = \pi \delta(x).
\end{equation}
This gives
\begin{align}
    \mathcal{I}_1 =  2\pi \int_{0}^{\infty} \frac{d T}{T^2} \big(G(T,T)+G(-T,-T)-2G(0,0)\big) - \frac{\pi^2}{L_\varphi}\frac{d^2}{dT^2}\left.G(T,T)\right|_{T=0},
\end{align}
The last term vanishes in the limit $L_\varphi \gg 1$ and we arrive at eq.\eqref{eq:one_step1}.

\section{Some integrals involving Airy functions}
\label{sec:app_b}
Consider the integral in eq.~\eqref{eq:G_cal_def}. From eqs.~\eqref{eq:G_func} and \eqref{eq:y_args} it follows that we have to evaluate integrals of the form
\begin{equation}\label{eq:tau_integrals}
\begin{split}
    &R_{0}(n) = \int_{-\infty}^{+\infty}d\tau \, \tau^n \Ai^2(x+b\,\tau^2), \\
    &R_{1}(n) = \int_{-\infty}^{+\infty}d\tau \, \tau^n \Ai(x+b\,\tau^2)\Ai'(x+b\,\tau^2), \\
    &R_2(n) = \int_{-\infty}^{+\infty}d\tau \, \tau^n \Ai'^2(x+b\,\tau^2),
\end{split} 
\end{equation}
for $n = 0,\,1,\,2$. Obviously, $R_0(n), \,R_1(n)$ and $R_2(n)$ all vanish for odd $n$. For even $n$
\begin{equation}
    R_0(n) = 2\int_0^{\infty} d\tau \, \tau^{n} \Ai^2(x+b\,\tau^2) = b^{-\frac{n+1}{2}}\int_0^{\infty} dt \, t^{\frac{n-1}{2}} \Ai^2(x+t),
\end{equation}
where we have substituted $t = b\,\tau^2$. Since Airy functions satisfy the equation
\begin{equation}\label{eq:airy_eq}
    \Ai''(z) - z\Ai(z) = 0,
\end{equation}
it follows that
\begin{equation}
    \left[\frac{d^2}{dx^2}-x\right]\Ai(x+t) = t \Ai(x+t).
\end{equation}
Integrating by parts and using this relation, we can establish the recurrent formula (see \cite{vallee2004airy})
\begin{equation}\label{eq:Ai2_integral_recurrent}
    \int_0^\infty dt \, t^m \Ai^2(x+t) = \frac{m}{2m+1}\left[\frac{1}{2}\frac{d^2}{dx^2}-2x\right]\int_0^\infty dt\,t^{m-1}\Ai^2(t+x),
\end{equation}
valid for $m>0$. For $m = -1/2$ there is a well-known result \cite{vallee2004airy}
\begin{equation}\label{eq:Ai2_integral_minus_half}
    \int_0^\infty d t \, t^{-\frac{1}{2}}\Ai^2(x+t) = \frac{1}{2}\Ai_1(2^{\frac{2}{3}}x).
\end{equation}
Using eqs.~\eqref{eq:Ai2_integral_recurrent} and \eqref{eq:Ai2_integral_minus_half} we can evaluate
\begin{align}
    R_0(0) =& \frac{1}{2}\Ai_1(2^{\frac{2}{3}}x)\,b^{-\frac{1}{2}}, \\
    R_0(2) =&  -\frac{1}{8} \left(2 x \text{Ai}_1(2^{\frac{2}{3}}  x)+2^{\frac{1}{3}} \text{Ai}'(2^{\frac{2}{3}}  x)\right)\, b^{-\frac{3}{2}}.
\end{align}
It is not hard to relate $R_1(n)$ to $R_0(n)$. First,
\begin{equation}
    R_1(n) = \frac{1}{2}\frac{d}{dx}R_0(n).
\end{equation}
Then it follows that
\begin{align}
    &R_1(0) = -2^{-\frac{4}{3}}\Ai(2^{\frac{2}{3}}x)\,b^{-\frac{1}{2}}, \\
    &R_1(2) = -\frac{1}{8}\Ai_1(2^{\frac{2}{3}}x)\,b^{-\frac{3}{2}}.
\end{align}
Note that
\begin{equation}
    \left[\frac{1}{2}\frac{d^2}{dx^2}-x\right]\Ai^2(x+t) = t \Ai^2(x+t) + \Ai'^2(x+t).
\end{equation}
This equation allows to relate $R_2(n)$ to $R_0(n)$
\begin{align}
    R_2(n) & = b^{-\frac{n+1}{2}}\int_0^{\infty} dt \, t^{\frac{n-1}{2}} \Ai'^2(x+t) = \nonumber \\
    & =  b^{-\frac{n+1}{2}}\left[\frac{1}{2}\frac{d^2}{dx^2}-x\right]\int_0^{\infty} dt \, t^{\frac{n-1}{2}} \Ai^2(x+t) - b^{-\frac{n+1}{2}}\int_0^{\infty} dt \, t^{\frac{n+1}{2}} \Ai^2(x+t) \nonumber \\
    &= \frac{1}{n+2}\left[\frac{n+3}{4}\frac{d^2}{dx^2}-x\right]R_0(n),
\end{align}
where in the last step we have used eq.~\eqref{eq:Ai2_integral_recurrent}. Using this relation, we finally evaluate
\begin{align}
    R_2(0) &= -\frac{1}{8} \left(2 x \, \text{Ai}_1(2^{\frac{2}{3}} x)+3 \cdot2^{\frac{1}{3}} \text{Ai}'(2^{\frac{2}{3}} x)\right)\,b^{-\frac{1}{2}}, \\
    R_2(2) &= \frac{1}{64} \left(2 x \left(2 x \, \text{Ai}_1(2^{\frac{2}{3}} x)+2^{\frac{1}{3}} \text{Ai}'(2^{\frac{2}{3}} x)\right)+5\cdot 2^{\frac{2}{3}} \text{Ai}(2^{\frac{2}{3}} x)\right)\,b^{-\frac{3}{2}}.
\end{align}

\section{Integrals over phase involved in dressed vertices}
\label{sec:app_c}
The coefficients in eq.~\eqref{eq:Gamma_cn} are given by the integrals
\begin{equation}\label{eq:Cn_functions_definition}
    C_n (s;p',p) = \int_{-\infty}^{+\infty}\frac{d\varphi}{2\pi}\, \varphi^n \exp\left(i s \varphi-i\frac{\alpha_{p'p}}{2}\varphi^2 +i\frac{4\beta_{p'p}}{3}\varphi^3\right),
\end{equation}
where
\begin{equation}\label{eq:alpha_beta_definitions}
    \alpha_{p'p} = e\left(\frac{pa}{kp} - \frac{p'a}{kp'}\right),\quad \beta_{p'p}=\frac{e^2 a^2}{8}\left(\frac{1}{kp}-\frac{1}{kp'}\right),
\end{equation}
which are expressed in terms of Airy functions as follows
\begin{align}
    &C_0(s;p',p) = \frac{1}{(4\beta_{p'p})^{1/3}}\Ai\big(y_{p'p}(s)\big) \, e^{i\theta_{p'p}(s)}, \label{eq:C0_function} \\
    &C_1(s;p',p) = \frac{1}{(4\beta_{p'p})^{1/3}} \left[ \frac{\alpha_{p'p}}{8\beta_{p'p}} \, \Ai\big(y_{p'p}(s)\big) - \frac{i}{(4\beta_{p'p})^{1/3} }\, \Ai'\big(y_{p'p}(s)\big) \right] \, e^{i\theta_{p'p}(s)}, \label{eq:C1_function} \\
    &C_2(s;p',p) = \frac{1}{(4\beta_{p'p})^{1/3}} \Bigg[ \left(\left(\frac{\alpha_{p'p}}{8\beta_{p'p}}\right)^2-\frac{y_{p'p}(s)}{(4\beta_{p'p})^{2/3}}\right) \Ai\big(y_{p'p}(s)\big)  \nonumber \\
    & \quad\quad\quad\quad\quad\quad - \frac{i\alpha}{(4\beta_{p'p})^{4/3} }\, \Ai'\big(y_{p'p}(s)\big) \Bigg]e^{i\theta_{p'p}(s)}, \label{eq:C2_function}
\end{align}
where
\begin{equation}\label{eq:Cn_substitutions}
    \theta_{p'p}(s) = \frac{\alpha_{p'p}}{8\beta_{p'p}}s-\frac{\alpha_{p'p}^3}{192\beta_{p'p}^2}, \quad y_{p'p}(s) = (4\beta_{p'p})^{2/3}\left(\frac{s}{4\beta_{p'p}} - \left(\frac{\alpha_{p'p}}{8\beta_{p'p}}\right)^2\right).
\end{equation}

\section{Verification of the two-step contribution to the trident process}
In this section, we  compare the two-step contribution to the trident process rate, namely, the first term in eq.~\eqref{eq:rate_direct}, or, equivalently, the rate given by eq.~\eqref{eq:W_trident_from_loops_2st} against (i) combination of two independently computed first-order process rates, (ii) the result of Ritus \cite{ritus1972vacuum}, and (iii) the result of King et al \cite{king2013trident}.

\subsection{Direct product of two first-order subprocesses}
By definition, the two-step rate is a combination of consecutive subprocesses of proton emission and pair production  \cite{ritus1972vacuum}:
\begin{equation}\label{eq:two_step_definition}
    W^{\mathrm{2\,st}}(\chi)=\frac{1}{2} \sum\limits_{\lambda} \int_0^\infty du \frac{dW_{\mathrm{rad},\lambda}(\chi,u)}{du}\times T_\gamma W_{\mathrm{cr},\lambda}(\chi_l).
\end{equation}
In this expression, factor $1/2$ accounts for the symmetry between the two final electron phase spaces [namely, $N_X=2$, see eq.~\eqref{eq:measure_Pi_tilde}]. Equation~\eqref{eq:two_step_definition} is a convolution of the differential rate $dW_{\mathrm{rad},\lambda}(\chi,u)/du$ for an electron with $p^0$ and $\chi$ to emit a photon with $\chi_l=u\chi/(1+u)$ and in polarization state $\varepsilon_\lambda$ with the total rate $W_{\mathrm{cr},\lambda}(\chi_l)$ of pair creation by this photon in a CCF. We imply that $W^{\mathrm{2\,st}}$ is averaged over the spin states of the initial electron and is summed over the final electrons and positron states. Furthermore, we also sum over two independent  polarization states of the intermediate photon. The first-order subprocess rates entering eq.~\eqref{eq:two_step_definition}  resolved in photon polarization  read \cite{king2013trident,king2013photon}:\footnote{In ref.~\cite{king2013photon}, the differential rate of photon emission $R_\gamma$ is normalized such that it coincides with the total rate of emission (see p. 559, eq. (49) in \cite{ritus1985quantum}) when \textit{averaged} over the final particle states. We normalize eq.~\eqref{dWrad}, such that the total rate is obtained by \textit{summing} over the final photon states. For this reason, our expression for $dW_{\mathrm{rad},\lambda}(\chi,u)/du$ has an additional factor $1/2$ as compared to $R_\gamma$.}
\begin{equation}\label{dWrad}
    \frac{dW_{\mathrm{rad},\lambda}(\chi,u)}{du}=-\frac{\alpha m^2}{2 p^0 (1+u)^2}\left[\Ai_1(z_1)+\frac{1}{z_1}\left(1+\frac{u^2}{1+u}+2\cos^2\phi\right)\Ai'(z_1)\right],
\end{equation}
\begin{equation}
    W_{\mathrm{cr},\lambda}(\chi_l)=2\frac{\alpha m^2}{l^0}\int_4^\infty \frac{dv}{v^{3/2}\sqrt{v-4}}\left[\Ai_1(z_2)+\frac{1}{z_2}\left(1-v+2\cos^2\phi\right)\Ai'(z_2)\right],
\end{equation}
where, $z_{1,2}$ are given in eq.~\eqref{eq:z1_z2_def} and $\phi$ is the angle of photon polarization. In the sum over the polarization states in eq.~\eqref{eq:two_step_definition}, we take $\phi=0$ and $\pi/2$. 

The large time factor $T$ in eq.~\eqref{eq:two_step_definition} corresponds to the time interval of the photon traverse in between emission and pair creation. Following Ritus \cite{ritus1972vacuum}, we associate $T$ to the large factor $\delta(\lambda=0)$ appearing eq.~\eqref{Wtr_from_im_M}, and express it via the large phase interval, $T_\gamma/l^0\mapsto L_\varphi/(kl)=\xi L_\varphi/m^2\chi\times(1+u)/u$, see also eq.~\eqref{eq:delta_zero}. The resulting expression for the two-step rates reads:
\begin{equation}\label{eq:W_two_step_appendix}
    \begin{split}
    W^{\mathrm{2\,st}}(\chi)=\frac{\alpha^2 m^2}{ p^0 \chi} & \xi L_\varphi  \int_{0}^{\infty}  du \int_4^\infty dv \frac{1}{u(1+u)v^{3/2}\sqrt{v-4}}\\
    &\times \left\lbrace \chi^{4/3}\frac{u^2(v-2)+(1+u)(2v-5)}{(1+u)^{5/3}v^{2/3}} \Ai'(t)\Ai'(\zeta) \right. \\
       &\quad\quad + \left[\frac{\chi u}{(1+u)v}\right]^{2/3}(v-2)\Ai_1(t)\Ai'(\zeta) \\
       &\quad\quad \left. - \left(\frac{\chi}{u}\right)^{2/3}\frac{u^2+2u+2}{1+u}\Ai'(t)\Ai_1(\zeta)-\Ai_1(t)\Ai_1(\zeta)\right\rbrace.
    \end{split}
\end{equation}
The definition given in eq.~\eqref{eq:W_trident_from_loops_2st} corresponds to the total two-step contribution, namely, $W^{\mathrm{2\,st}}=W^{\mathrm{2\,st}}_{\mathrm{dir}}+W^{\mathrm{2\,st}}_{\mathrm{ex}}$. Hence, $W^{\mathrm{2\,st}}_{\mathrm{dir}}=W^{\mathrm{2\,st}}/2$ coincides with the term proportional to $\xi L_\varphi$ in eq.~\eqref{eq:rate_direct} obtained in the tree-level calculation, as well as with eq.~\eqref{eq:W_trident_from_loops_2st} extracted from the two-loop result.

\subsection{Comparison against the result of Ritus}
Let us consider the two-step contribution resolved in the initial electron spin states, that we obtained from the two-loop result, namely, eqs.~\eqref{eq:W_trident_from_loops_2st} and \eqref{eq:phi}. To compare it against the corresponding formula in ref.~\cite{ritus1972vacuum}, we rewrite these expressions by integrating by parts. 

Let us use the equality
\begin{equation}
    \int_4^\infty dv\,f(v)\Ai_1(z_2)=\int_4^\infty dv \Ai'(z_2)\left(\frac{\chi_l}{v}\right)^{2/3}\frac{2F(v)-3vf(v)}{3v^{5/3}},
\end{equation}
where $F(v)=\int dv f(v)$. It is obtained by integrating by parts twice and using eq.~\eqref{eq:airy_eq}. By taking $f(v)=1/v^{3/2}\sqrt{v-4}$ [hence, $F(v)=\sqrt{v-4}/2\sqrt{v}$], we can use this expression to rewrite terms proportional to $\Ai_1(z_2)$ in eq.~\eqref{eq:phi} (at $\lambda=0$), which gives:
\begin{equation}\label{eq:W_two_step_appendix_Ritus}
    \begin{split}
    W^{\mathrm{2\,st}}_s(\chi)&= W^{\mathrm{2\,st}}_{\mathrm{dir},s}(\chi)+W^{\mathrm{2\,st}}_{\mathrm{ex},s}(\chi)\\
    &=\frac{2 \alpha^2 m^2}{p^0 \chi}    \int_{0}^{\infty}  \frac{du}{(1+u)^2} \int_4^\infty \frac{dv}{v^{3/2}\sqrt{v-4}}\, 2\pi \left[\frac{\xi L_\varphi}{4\pi \chi} \frac{1+u}{u} \right] \frac{2v+1}{3 z_2}\Ai'(z_2)\\
    &\quad\quad\quad\quad\quad\quad\quad\quad\quad \times \left\lbrace \Ai_1(z_1) + \frac{1}{z_1}\left[\frac{u^2+2u+2}{1+u}-\frac{3}{2v+1}\right]\Ai'(z_1)\right. \\
    &\quad\quad\quad\quad\quad\quad\quad\quad\quad\quad\quad\quad\quad\quad\quad\quad\quad \left. + 2\gamma_s z_1 \frac{3u-2v+2}{(1+u)(2v+1)}\Ai(z_1) \right\rbrace.
    \end{split}
\end{equation}
Notice that the combination in the square brackets containing $\xi L_\varphi$ is the regularized representation of  $\delta(\lambda/m^2=0)$, see eq.~\eqref{eq:delta_zero}. Equation \eqref{eq:W_two_step_appendix_Ritus} coincides with the $\delta(0)$ term interpreted by Ritus as the two-step contribution, c.f. eq. (14) in ref.~\cite{ritus1972vacuum}. Note that the overall minus sign for this term appearing therein is attributed to the subtraction of the two-step contribution from the polarization correction to the photon emission process.

\subsection{Comparison against the result of King et al}
Study \cite{king2013trident} reported a calculation that is analogous to ours in section \ref{sec:vi}. As notations and intermediate steps in these two derivations differ, it is worth comparing the results. While we do not find  explicit expressions for the one-step term in ref.~\cite{king2013trident}, eqs.~(16)-(17) therein provide the explicit probability $P^{(2)}$ for the two-step process  and therefore can be compared with our results. First, we pass to the rate $W^{(2)}=P^{(2)}/T$, where $T$ is the large time attributed to the whole process. Following ref.~\cite{king2013trident} [see the text after eq.~(A20) therein], we relate $T/p^0\sim L_\varphi/(kp)$, where  $L_\varphi$ is defined as in eq.~\eqref{eq:L_phi_def}. Note that $L_\varphi$ corresponds to $\Delta\varphi$ in the notation of ref.~\cite{king2013trident}. To this end, we write $W^{(2)}=(m^2/2p^0) \xi L_\varphi \chi \mathcal{I}^{(2)}$, where $\mathcal{I}^{(2)}$ is given in ref.~\cite{king2013trident}, see eqs.~(16)-(17) therein. Finally, we translate integrals appearing in $\mathcal{I}^{(2)}$ to our notations by using the following substitutions:
\begin{equation}
  \chi_1=\chi,\quad \chi_2=\frac{\chi_l}{1+u},\quad \chi_3=\frac{u\chi_l}{1+u}\left(1+\sigma \sqrt{1-\frac{4}{v}}\right), 
\end{equation}
with the Jacobian determinant $|J|=u\chi^2/(1+u)^3 v^{3/2}\sqrt{v-4}$, so that after the substitution the integrand should be summed over the branches $\sigma=\pm 1$. When rewritten in terms of variables $0<u<\infty$, $4<v<\infty$, the $\theta$-function entering eq.~(16) in ref.~\cite{king2013trident} becomes trivial, as its argument becomes positive (for $\chi>0$). As a result, we recover our eq.~\eqref{eq:W_trident_from_loops_2st} up to an extra numerical factor $(-2)$. According to ref.~\cite{king2013trident}, $P^{(2)}$ is associated with the total probability, that is, a sum of the direct and exchange terms, which explains the factor of $2$. The overall minus sign that we obtained is probably due to a typo.


\providecommand{\href}[2]{#2}\begingroup\raggedright\endgroup

\end{document}